%% file: main.tex
\documentclass{article}

\newif\ifanonymous
\anonymoustrue

\usepackage{appendix}
\usepackage{placeins}
\PassOptionsToPackage{numbers,sort&compress}{natbib}

\usepackage[preprint]{neurips_2025} 

\usepackage{authblk} 
\usepackage{amsmath}
\usepackage{units}
\usepackage{xspace}

\newcommand{\ates}{\ensuremath{\alpha_{\rm tes}}\xspace}
\newcommand{\ajes}{\ensuremath{\alpha_{\rm jes}}\xspace}
\newcommand{\asoftmet}{\ensuremath{\alpha_{\rm soft\_met}}\xspace}
\newcommand{\attbar}{\ensuremath{\alpha_{\rm ttbar\_scale}}\xspace}
\newcommand{\adiboson}{\ensuremath{\alpha_{\rm diboson\_scale}}\xspace}
\newcommand{\abkg}{\ensuremath{\alpha_{\rm bkg\_scale}}\xspace}

\usepackage[utf8]{inputenc} 
\usepackage[T1]{fontenc}    
\usepackage{url}            
\usepackage{booktabs}       
\usepackage{amsfonts}       
\usepackage{nicefrac}       
\usepackage{microtype}      
\usepackage{xcolor}         
\usepackage{ulem}
\usepackage{adjustbox}      
\usepackage{subcaption}

\usepackage{hyperref}   
\hypersetup{
  colorlinks   = true, 
  urlcolor     = purple, 
  linkcolor    = blue, 
  citecolor    = red 
}

\usepackage[disable]{todonotes}  

\newcommand{\instructions}[1]{\todo[inline,color=yellow!40]{Instructions: #1}}

\newcommand{\david}[1]{\todo[inline,color=green!40]{#1 \\-- David}}

\begin{document}

\title{FAIR Universe HiggsML Uncertainty \\ Dataset and Competition}

\author[8]{Lisa Benato} 
\author[1]{Wahid Bhimji}
\author[1]{Paolo Calafiura} 
\author[2,7]{Ragansu Chakkappai}
\author[1] {Po-Wen Chang}
\author[3]{Yuan-Tang Chou}
\author[1]{Sascha Diefenbacher}
\author[1,4]{Jordan Dudley}
\author[5]{Ibrahim Elsharkawy}
\author[1]{Steven Farrell}
\author[1,5]{Aishik Ghosh}
\author[8]{Cristina Giordano}
\author[7]{Isabelle Guyon}
\author[1]{Chris Harris}
\author[9]{Yota Hashizume}
\author[3]{Shih-Chieh Hsu}
\author[1,3,10]{Elham E Khoda}
\author[8]{Claudius Krause}
\author[8]{Ang Li}
\author[1]{Benjamin Nachman}
\author[1]{Peter Nugent} 
\author[2,7]{David Rousseau}
\author[8]{Robert Schoefbeck}
\author[8]{Maryam Shooshtari}
\author[8]{Dennis Schwarz}
\author[1]{Benjamin Thorne} 
\author[7]{Ihsan Ullah}
\author[8]{Daohan Wang}
\author[3]{Yulei Zhang}

\affil[1]{Lawrence Berkeley National Laboratory} 
\affil[2]{Universit\'e Paris-Saclay, CNRS/IN2P3, IJCLab}
\affil[3]{University of Washington, Seattle}
\affil[4]{University of California, Berkeley}
\affil[5]{University of Illinois Urbana-Champaign}
\affil[6]{University of California, Irvine}
\affil[7]{ChaLearn}
\affil[8]{Institute for High Energy Physics, Vienna}
\affil[9]{Kyoto University}
\affil[10]{University of California, San Diego}




\maketitle



\begin{abstract}

The FAIR Universe – HiggsML Uncertainty Challenge focused on measuring the physical properties of elementary particles with imperfect simulators. Participants were required to compute and report confidence intervals for a parameter of interest regarding the Higgs boson while accounting for various systematic (epistemic) uncertainties. The dataset is a tabular dataset of 28 features and 280 million instances. Each instance represents a simulated proton-proton collision as observed at CERN's Large Hadron Collider in Geneva, Switzerland. The features of these simulations were chosen to capture key characteristics of different types of particles. These include primary attributes, such as the energy and three-dimensional momentum of the particles, as well as derived attributes, which are calculated from the primary ones using domain-specific knowledge. Additionally, a label feature designates each instance's type of proton-proton collision, distinguishing the Higgs boson events of interest from three background sources. As outlined in this paper, the permanent release of the dataset allows long-term benchmarking of new techniques. The leading submissions, including Contrastive Normalising Flows and Density Ratios estimation through classification, are described. Our challenge has brought together the physics and machine learning communities to advance our understanding and methodologies in handling systematic uncertainties within AI techniques.

\end{abstract}


\instructions{
To be submitted in the NeurIPS2025 Dataset and benchmark track general instructions https://neurips.cc/Conferences/2025/CallForPapers specified  https://neurips.cc/Conferences/2025/CallForDatasetsBenchmarks .
Deadline for abstract 11th May AoE, full paper 15th may, technical appendices, 22 May\\
Author list to be updated to have all HEPHY members, Ibrahime and hzume. 
Page allocation 9 pages, then references, checklist, and the optional technical appendices
Very important : this paper will be evaluated by Computer Scientist who have mostly no training in physics !

}


\section{Introduction}

\subsection{Background and impact}
\label{sec_background}

For several decades, the discovery space in almost all branches of science has been accelerated
dramatically due to increased data collection brought on by the development of larger, faster
instruments. More recently, progress has been further accelerated by the emergence of powerful AI approaches, including deep learning, to exploit this data. However, an unsolved challenge that remains, and \textit{must} be tackled for future discovery, is how to effectively quantify and reduce uncertainties, including understanding and controlling \textit{systematic} uncertainties (also named \textit{epistemic} uncertainties in other fields). 
A compelling example is found in analyses to further our fundamental understanding of the universe through analysis of the vast volumes of particle physics data produced at CERN, in the Large Hadron Collider (LHC)~\cite{Evans_2008}.
Ten years ago, part of our team co-organised the Higgs Boson Machine Learning Challenge (HiggsML~\cite{pmlr-v42-cowa14,kaggle_higgsml2014}, the most popular Kaggle challenge at the time attracting 1785 teams. This challenge has significantly heightened interest in applying Machine Learning (ML) techniques within High-Energy Physics (HEP) and, conversely, has exposed physics issues to the ML community. Whereas previously, the most effective methods predominantly relied on boosted decision trees, Deep Learning has since gained prominence (see, e.g., HEP ML living review~\cite{hepmllivingreview}). While the LHC has not (yet) discovered new physics beyond the Higgs boson, it has accumulated vast data and will continue to accumulate more data well into the next decade. There is a discovery potential in very precise measurements of particle properties, particularly of the Higgs boson.

High-energy physics relies on statistical analysis of aggregated observations. Therefore, the interest in uncertainty-aware ML methods in HEP is nearly as old as the application of ML in the field. Advanced efforts that integrate uncertainties into the ML training include approaches that explicitly depend on nuisance parameters~\cite{Cranmer:2015bka,Baldi:2016fzo,Brehmer:2019xox,Brehmer:2018hga,Brehmer:2018kdj,Brehmer:2018eca,Nachman:2019dol,Ghosh:2021roe,Rozet:2021diu,ATLAS:2024ynn}, that are insensitive to nuisance parameters~\cite{Blance:2019ibf,Englert:2018cfo,Louppe:2016ylz,Dolen:2016kst,Moult:2017okx,Stevens:2013dya,Shimmin:2017mfk,Bradshaw:2019ipy,ATL-PHYS-PUB-2018-014,DiscoFever,Wunsch:2019qbo,Rogozhnikov:2014zea,10.1088/2632-2153/ab9023,clavijo2020adversarial,Kasieczka:2020pil,Kitouni:2020xgb,Estrade:2019gzk,Ghosh:2021hrh}, that use downstream test statistics in the initial training~\cite{Wunsch:2020iuh,CMS:2025cwy,Heinrich:2022qlq,Elwood:2020pik,Xia:2018kgd,DeCastro:2018psv,Charnock_2018,Alsing:2019dvb,Simpson:2022suz,Feichtinger:2021uff,Layer:2023lwi}, and that use Bayesian neural networks for estimating uncertainties~\cite{Kasieczka:2020vlh,Bollweg:2019skg,Araz:2021wqm,Bellagente:2021yyh}.  Many of these topics were covered in recent forward-looking review-type articles in Refs.~\cite{Dorigo:2020ldg, Chen:2022pzc}.
%
However, these developments all report technique performance on different ad-hoc datasets, so it is difficult to compare their merits. The Fair Universe HiggsML Uncertainty Challenge, an official NeurIPS 2024 competition, aimed to provide a common ground, with a dataset of sufficient complexity, equipped with systematic bias parameterisations, and a metric.

We aim to address the issue of systematic uncertainties within a specific domain. Yet, the techniques developed by the challenge participants will apply to identifying, quantifying, and correcting systematic uncertainties in other areas, particularly other science disciplines.

\subsection{Novelty}
\label{sec_novelty}
This entirely new public competition has built on our experience running several competitions in particle physics and beyond. These include the original HiggsML challenge \cite{pmlr-v42-cowa14}, the TrackML Challenges (NeurIPS 2018 competition) \cite{TrackMLAccuracy2019,TrackMLThroughput}, the LHC Olympics \cite{lhc-olympics}, AutoML/AutoDL \cite{automl, autodl}, and other competitions. 
Building on the foundation of the HiggsML challenge, this competition introduces a significant change by using simulated data that includes biases (or {\it systematic effects}). 
In addition, participants were asked to provide a confidence interval and not just a point estimate. 

While there have been previous challenges focusing on meta-learning and transfer-learning, such as the NeurIPS 2021 and 2022 meta-learning challenges~\cite{elbaz2021pmlr, carriónojeda2022neurips22}, Unsupervised and Transfer Learning~\cite{guyon2012analysis}, challenges related to bias e.g. Crowd bias challenge \cite{crowd-bias-challenge}, and those addressing distribution shifts, like the Shifts challenge\cite{malinin2022shifts20extendingdataset} series, and CCAI@UNICT 2023 \cite{ccaiunict-2023}, this is the first challenge and dataset that requires participants to handle systematic uncertainty. 
Moreover, this project is connecting the Perlmutter system at NERSC~\cite{nersc}, a large-scale supercomputing resource featuring over 7000 NVIDIA A100 GPUs, with Codabench~\cite{codabench}, a new version of the renowned open-source benchmark platform Codalab~\cite{carlens2025state,codalab_competitions_JMLR}. 
Due to its complexity, the process of generating events was computationally intensive; use of the Perlmutter supercomputer allowed us to create a vast amount of data -- hundreds of millions of events compared to less than a million events for the HiggsML competition, which will serve as a long-lasting benchmark.

\section{Data}

The dataset is publicly available on the Zenodo platform~\cite{https://doi.org/10.5281/zenodo.15131565}. 
The data is saved as a tabular parquet~\cite{Vohra2016} file of 16~GB and is accompanied by a Croissant JSON metadata file. The dataset comprises 280M simulated proton-proton collision events and is weighted to represent two weeks of LHC data taking. A separate 120M i.i.d dataset has been used for the final results in \autoref{sec:results} and is kept private for future over-training checks.

We are using a simulated particle physics dataset for this competition to produce data representative of high-energy proton collision data collected by the ATLAS experiment~\cite{2008atlas} at the LHC. The dataset~\cite{https://doi.org/10.5281/zenodo.15131565} was created with two widely-used simulation tools, Pythia 8.2~\cite{Sjostrand:2014zea} and Delphes~3.5.0 \cite{deFavereau:2013fsa}; all the configuration and data pre-selection code is available from~\cite{FAIR_Universe_dataGen}. This required 1.8 million CPU core hours. 
We have organised the dataset into a tabular format where each row corresponds to a collision event and each of the 28 columns corresponds to a feature. The detailed dataset description is in \autoref{sec_collisions}, \autoref{sec_relativity} and \autoref{sec_data_detailed}; it is mostly taken from the public unpublished Fair Universe whitepaper~\cite{FairU_whitepaper} which served as detailed documentation for the competition.
Part of the features are primary features, essentially the energy and direction of a small set of particles, and the remainder are derived features, computed from the primary ones with domain knowledge. 
The events are divided into two categories (see \autoref{tab_data_table}): signal and background. The signal category includes collision events with a Higgs boson decaying into pairs of tau particles (see \autoref{fig_particles_diag}) (one decaying, into, in addition to neutrino(s), a light lepton, the other one into a set of hadrons hence the name hadronic tau), while the background category includes other processes (subcategories) leading to a similar final state, but without an intermediate Higgs boson.

In addition, we provide a biasing script capable of manipulating a dataset by introducing six parameterised distortions as a function of six corresponding {\it Nuisance Parameters}\footnote{The name Nuisance Parameter, commonly used in the physics literature, refers to a parameter governing a specific parameterisation of a systematic bias. Nuisance Parameters can be in part constrained from the data itself. Still, the name implies that constraining them is only interesting as an auxiliary task in the process of determining a parameter of interest like the signal strength $\mu$.} (the systematic biases); see details in \autoref{sec_syst}. For example, a detector miscalibration can cause a bias in other features in a cascading way,  or in another case, the magnitude of a particular background (e.g. the $t\bar{t}$) contribution can change so that the feature distributions can be different. In both cases, the inference would be done on a dataset not i.i.d. to the training dataset.

\section{Tasks and application scenarios}

The participant's objective is to develop an estimator for the number of Higgs boson events in a dataset analogous to results from LHC experiments. Such a measurement is typical of those carried out at the LHC, which allows us to strengthen (or invalidate) our understanding of the fundamental laws of nature.

The primary metric is the {\it signal strength} ($\mu$), which is the number of estimated Higgs boson events divided by the number of such events predicted by the Standard Model, which is the reference theory.  The challenge involved estimating $\mu$'s true value, $\mu_{true}$, which may vary from one (in practice for the challenge in the range 0.1 to 3) and is inherently unknown. 

Participants were tasked with generating a 68.27\% Confidence Interval (CI) for $\mu$, incorporating both aleatoric (random) and epistemic (systematic) uncertainties rather than a single-point estimate. The six different systematic uncertainties are implemented in \autoref{sec_syst}. 

The primary simulation dataset assumes a $\mu$ of one. Participants receive a training subset, where events are labelled based on their event type (e.g. Higgs boson event).
 We provide a script to generate unlabelled pseudo-experiment datasets from the primary simulation dataset for any value of $\mu$ and the six systematic biases. The participant's model should be able to reverse the process and provide a 68.27\% CI on $\mu$ for any pseudo-experiment.

In a machine learning context, the task resembles a transduction problem with distribution shift: it requires constructing a $\mu$ interval estimator from labelled training data and biased unlabelled test data. One possibility is to train a classifier to distinguish Higgs boson from the background, with robustness against bias achieved possibly through data augmentation (or an adversarial approach, or black box optimisation or any other novel approach)  via the provided script.

This challenge shifts focus from the qualitative discovery of individual Higgs boson events (which was the focus of our first challenge~\cite{pmlr-v42-cowa14}) to the quantitative estimation of overall Higgs boson counts in test sets, akin to assessing disease impact on populations rather than diagnosing individual cases.

\subsection{Metrics}
\label{sec_metrics}

\begin{figure}[htb]
    \centering
    \begin{subfigure}[b]{0.48\textwidth}
        \includegraphics[width=\textwidth, height=5.5cm]{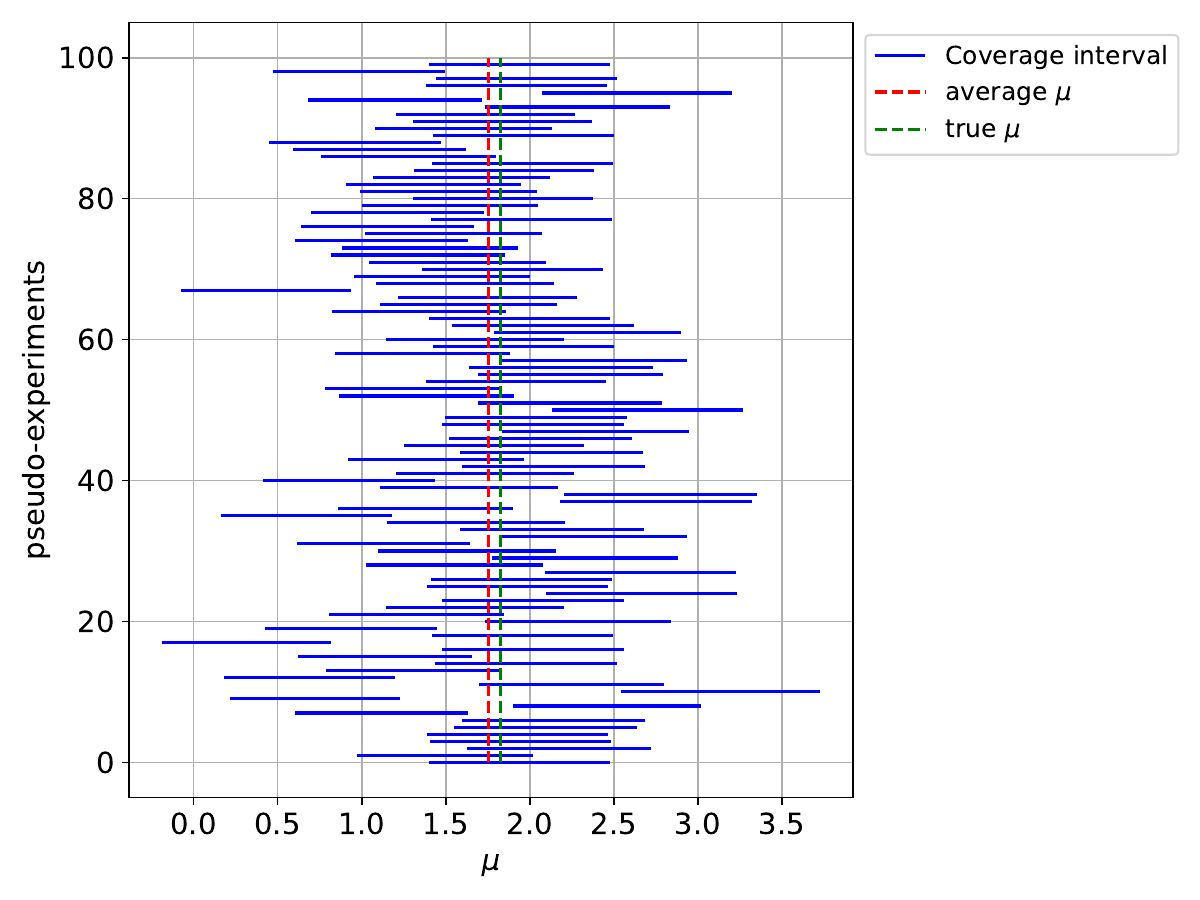}
        \caption{}
        \label{fig:coverage}
    \end{subfigure}
    \hfill 
    \begin{subfigure}[b]{0.48\textwidth}
        \includegraphics[width=\textwidth, height=5.5cm]{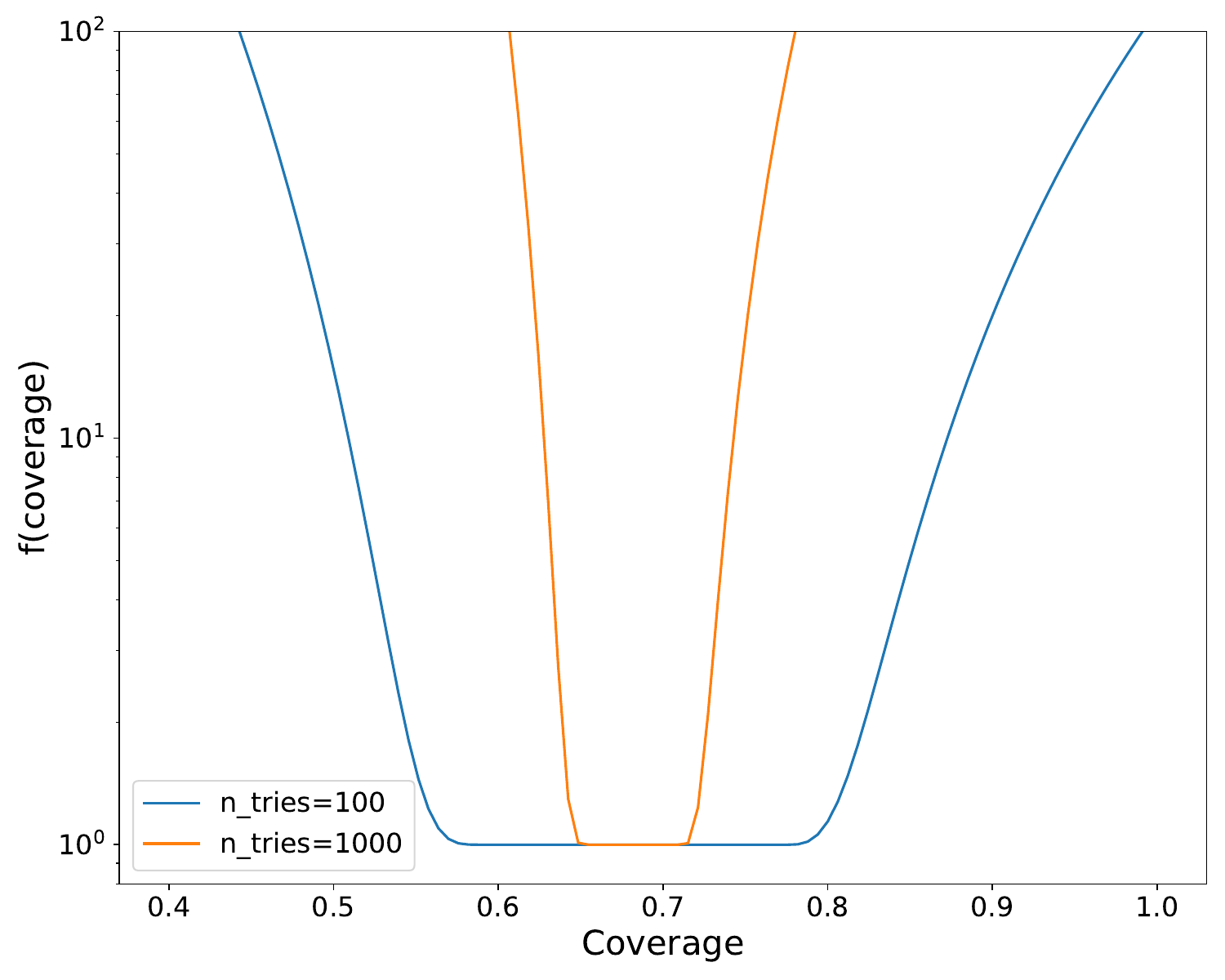}
        \caption{}
        \label{fig:pen_fun}
    \end{subfigure}
   \caption{ (\ref{fig:coverage}) \textit{Coverage plot}: all the predicted intervals (blue lines) for each pseudo experiment generated for a given $\mu_{\rm true}$ (vertical dotted line). The coverage  (here $70\pm5\%$) is determined by the fraction of time the horizontal blue lines intersects the vertical line. (\ref{fig:pen_fun}) Penalising function as a function of the coverage value $c$, for two values of $N_{\textrm test}$, the number of pseudo-experiments.}
\end{figure}

Participants provided a model that can analyse a pseudo-experiment to determine $(\mu_{16}, \mu_{84})$, the bounds of the 68.27\% (approximately one standard deviation of a standard normal distribution) Confidence Interval (CI) for $\mu$. The model is evaluated from the set of $[\mu_{16,i},\mu_{84,i}]$ intervals obtained from $N_{\text{test}}$ pseudo-experiments, see \autoref{fig:coverage}.  The model's performance is assessed based on two criteria: {\bf Average Interval Width} $w$ (the smaller the better) computed as $ w = \frac{1}{N_{test}} \sum_{i=1}^{N} |\mu_{84,i} - \mu_{16,i} |  $  and the {\bf Coverage}, the frequency with which $\mu_{\rm truth}$ is covered by the CI (the closer to the standard 68.27\% probability the better) computed as $c = \frac{1}{N_{test}} \sum_{i=1}^{N}\ 1\  \textrm{if} \  \mu_{\textrm{true}, i} \in [\mu_{16,i}, \mu_{84,i}]$.

A penalising function $f$ has been defined to penalise the departure of $c$ from the expected 68.27\%, taking into account 
$\sigma_{68}= \sqrt{\frac{(1-0.6827)0.6827 }{N_{\textrm{test}}}}$$ $ the binomial statistical error on $c$:

\begin{equation}
f(c)=1  + \mathbb{I}_{c  < 0.6827-2\sigma_{68}} \cdot \left|\frac{c-(0.6827-2\sigma_{68})}{\sigma_{68}}\right|^{4} + 
\mathbb{I}_{c  > 0.6827+2\sigma_{68}} \cdot \left|\frac{c-(0.6827+2\sigma_{68})}{\sigma_{68}}\right|^{3}\,.
\end{equation}

We opted for an asymmetric penalty function because, within the High Energy Physics (HEP) field, overestimating uncertainty is deemed more acceptable than underestimating it~\cite{Nachman:2012zf,Nachman:2014hsa}. Hence, coverage exceeding 68.27\%  incurs a lesser penalty than coverage falling below 68.27\%. The final {\bf Quantile Score} (the larger the better) used to rank participants is calculated as follows:
\begin{align}
    \textrm{score} = - \ln((w+\epsilon)f(c)),
\end{align}
where $w$ represents the average width of the Confidence Interval, $c$ is the coverage, and $\epsilon=10^{-2}$ is a regularisation term to guard against submissions that report unrealistically narrow CIs. 
To ensure efficient use of resources, each participant's model inference was executed across 100 pseudo-experiments times 10 trials, each with distinct values of $\mu_{\rm truth}$, with a time limit of 20s per inference on CPU or GPU. In the Final phase of the competition, each participant's best submission was evaluated over 100 pseudo-experiments, times 1000 trials, to minimize the statistical variance.

\subsection{Limitations}
\label{sec_limitations}

The main limitation of the setup is that biases can be exactly parameterised: we are in the "known unknowns" regime. "Unknown unknowns", unexpected biases, are not covered.  

The dataset has been produced using well-known standard software for event generation and detector simulation. However, a proper physics measurement would require more complex software, several orders of magnitude slower, yielding marginally different simulated data. The methods developed on our dataset would perform equally well, provided they are fully retrained.

For each instance of the datasets, the features provided are essentially the energy and direction of a small set of particles, and derived quantities. A real physics measurement may also rely on additional quantities related to the quality of particle identification or to other particles in the same proton-proton collision. Nevertheless, the algorithms developed on our dataset should require limited added complexity to deal with additional features. 

\section{Software}

Alongside the dataset, a GitHub repository~\cite{FAIR_Universe_dataset} with the relevant code for reading and analysing it is made available. This includes a Jupyter notebook starting kit, simple baseline models, a small sample of the dataset, and code to compute the score.

\begin{figure}[htb]
    \centering
    \begin{subfigure}[b]{0.48\textwidth}
        \includegraphics[width=\textwidth, height=5.5cm]{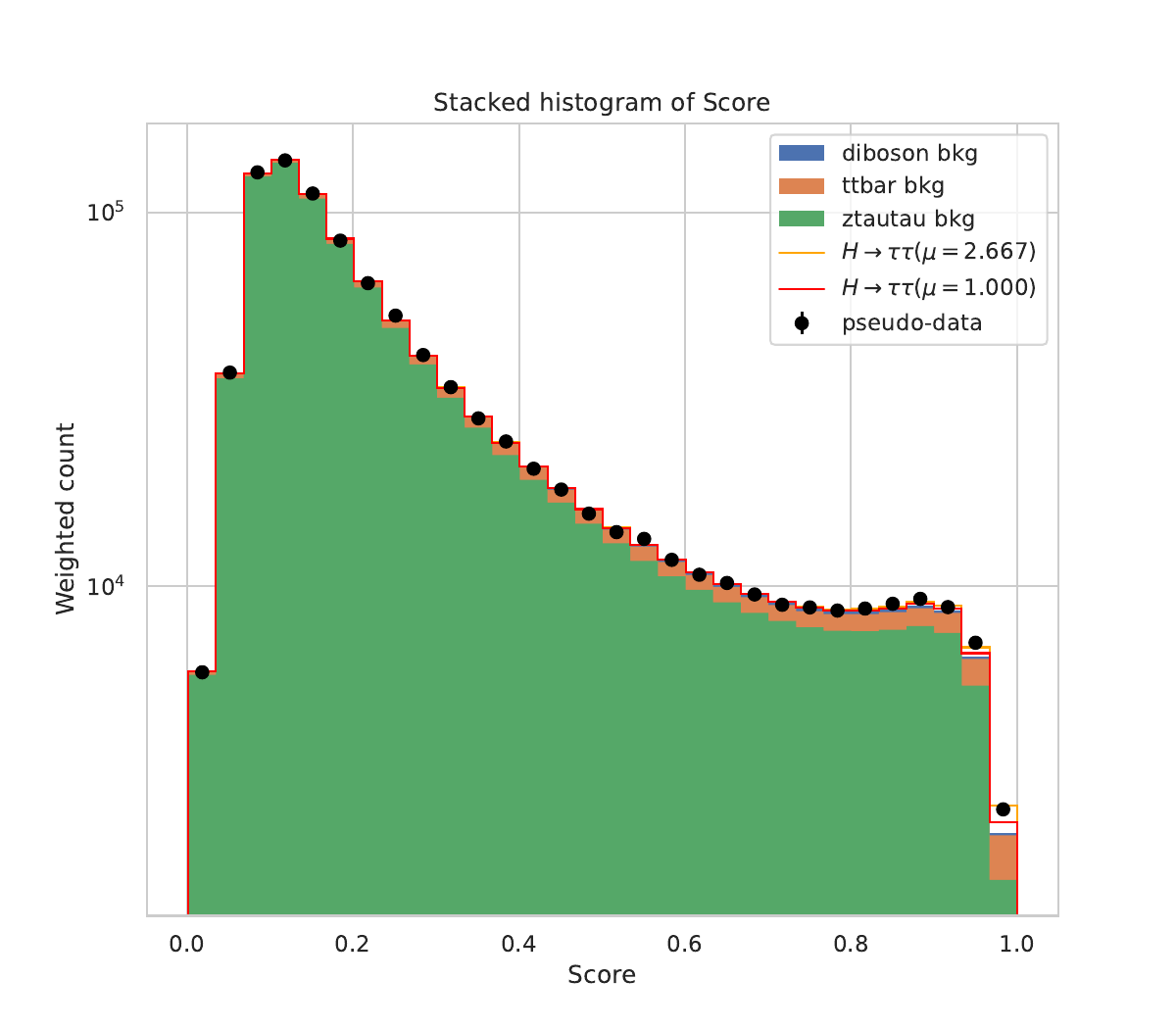} 
        \caption{}
        \label{fig:stacked_histogram}
    \end{subfigure}
    \hfill 
    \begin{subfigure}[b]{0.48\textwidth}
        \includegraphics[width=\textwidth, height=5.5cm]{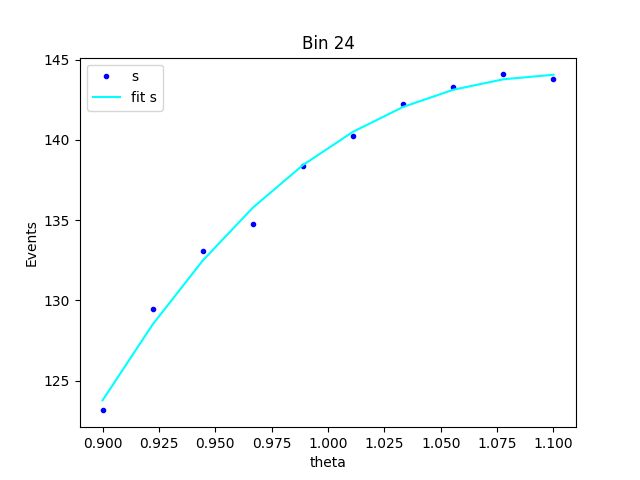}
        \caption{}
        \label{fig:NN_s_24}
    \end{subfigure}
    \caption{(a) classifier score for unlabelled test data (black points), and holdout data for (1) background events $Z\rightarrow \tau\tau$ (solid green), (2) background $t \bar{t}$ (solid orange) (3) background di-boson (solid blue, hardly visible) (4) signal events $H\rightarrow \tau\tau$ for $\mu=1$ (red line), and (5) signal events fitted histogram to test data, leading to estimated $\mu=2.667$ (orange line) (b) model of the bin content vs Nuisance Parameter $\theta$ for bin 24, as an example. \label{fig_score}}
\end{figure}

The \textbf{Starting Kit} kit includes code for installing necessary packages, loading and visualising data, training and evaluating a model with the metrics described in \autoref{sec_metrics}. The \textbf{Baseline} method estimates $\mu$ using standard techniques without directly addressing systematic uncertainties for simplicity. Initially, it utilises a classifier (based on an XGBoost Boosted Decision Tree) trained on a subset of the training data to enhance the signal event density and reduce the $\mu$ estimator variance. The classifier's decision threshold is fixed heuristically. $\mu$ is then estimated from these filtered events, assuming a Poisson distribution, enabling interval maximum likelihood estimation. Further refinement involves binning events based on their classifier score and estimating $\mu$ per bin. A holdout  dataset, is used to predict the amount of background and signal in each bin for $\mu=1$. This calibration step then permits estimating $\mu$ (and the corresponding CI) on each pseudo-experiment. On \autoref{fig:stacked_histogram},  the alignment of maximum likelihood estimation (orange line) with unlabelled data (black line) indicates the method success, in the absence of any bias.

When unknown biases occur, the prediction on the amount of background and signal events per bin will be wrong, biasing the estimation of $\mu$.
To address the problem of systematic errors, we use the holdout dataset with biases by different amounts of the Nuisance Parameter ($\theta$) and then build a calibration curve to estimate the signal and background in each bin. \autoref{fig:NN_s_24} shows one such fit curve for the 24th bin (just as an example). Now, instead of $\mu$ depending on $S$ and $B$, it will depend on fit functions $S(\theta)$ and $B(\theta)$. Finally, the minimisation function now regresses both $\mu$ and $\theta$, thus making the model less susceptible to systematic bias. But this is only limited to one nuisance parameter; participants are encouraged to enhance the Baseline model, for instance, by modifying the architecture or training protocol to improve resilience against biases, attempting to directly model the biases, or refining the estimator through a bias-aware model. 

Another way to see it is that, armed with the biasing script which can produce a dataset for any value of the six Nuisance Parameters and the signal strength $\mu$, the participants could train a model which could regress the seven parameters for any pseudo-experiment and report the Confidence Interval on $\mu$. This was actually done with different techniques by the winning trio (\autoref{sec:results}).




\section{Competition results and best submissions}
\label{sec:results}
At the end of the competition, a clear trio was at the top of the public leaderboard: HEPHY with a quantile score of 0.878, followed by Ibrahime (0.823) and Hzume (0.179).  All submissions have been reevaluated on a new dataset (i.i.d. to the original one). The evaluation was done on  1000 trials of 100 pseudo-experiments (each trial with a given value of $\mu$ randomised between 0.1 and 3), instead of 10 trials for the public leaderboard. All submissions were run on the same pseudo-experiments, instead of separate pseudo-experiments for the public leaderboard.  

\autoref{fig:three_scores} shows the results for all trials for the trio. The CI width is seen falling at large values of $\mu$: this is due to the clipping the Confidence Interval to a maximum value of~3 (which was not done in \autoref{fig:coverage}), which  was the maximum value in this competition. Such clipping would be meaningless in the context of a real physics measurement where $\mu$ is truly unknown. This is the only "hack" specific to the competition context that could be identified. As far as the score is concerned, HEPHY and Ibrahim are very close. When merging all trials, the scores obtained by the top trio are: HEPHY -0.582, Ibrahim -0.576 and HZUME -2.16.  An additional bootstrap analysis of the variance of these results showed that HEPHY and Ibrahim cannot be reliably ranked, hence the final rankings :

\begin{figure}[htb]
    \centering
    \begin{subfigure}[b]{0.32\textwidth}
        \includegraphics[width=\textwidth]{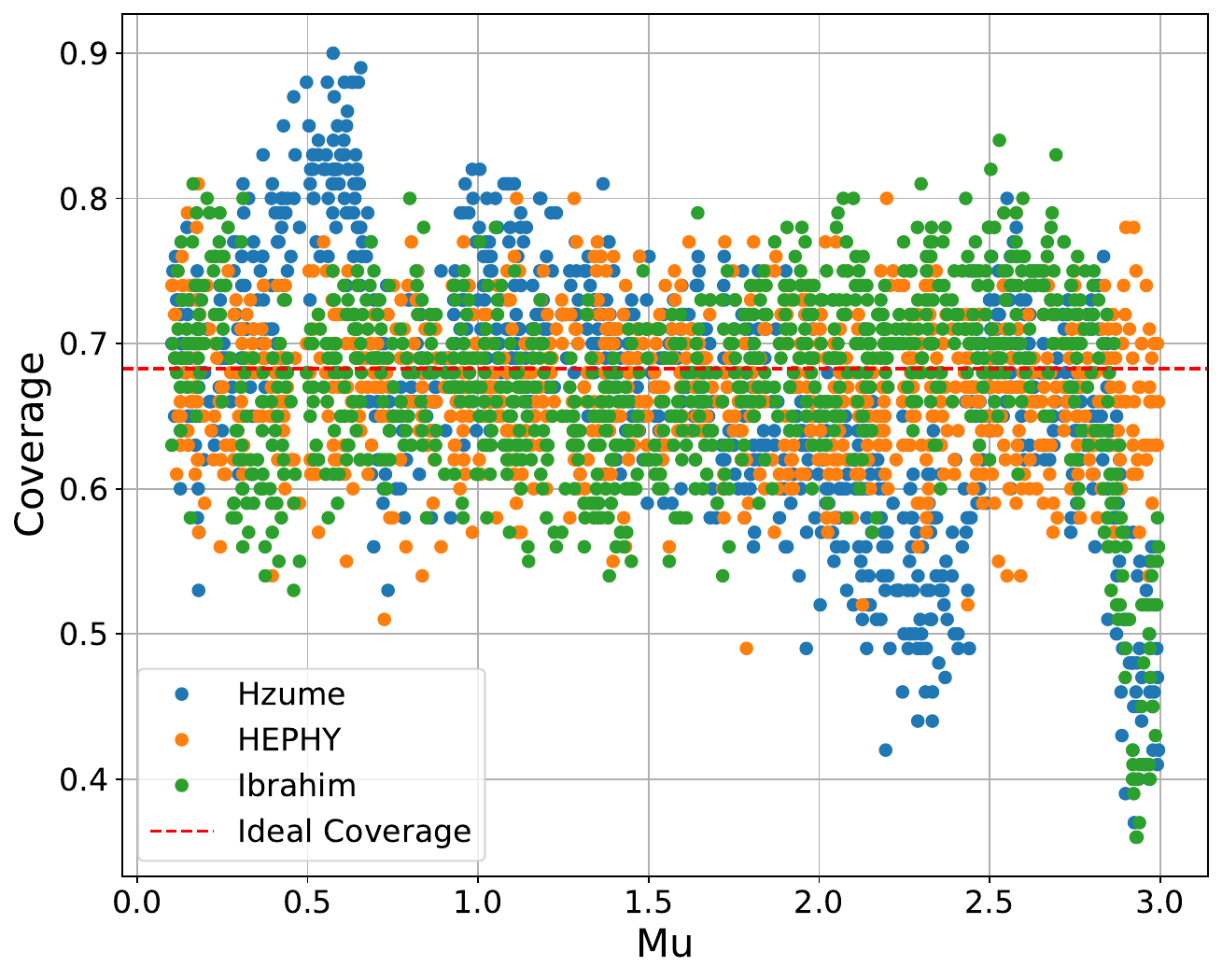}
        \caption{}
        \label{fig:coverage_vs_mu}
    \end{subfigure}
    \hfill 
    \begin{subfigure}[b]{0.32\textwidth}
        \includegraphics[width=\textwidth]{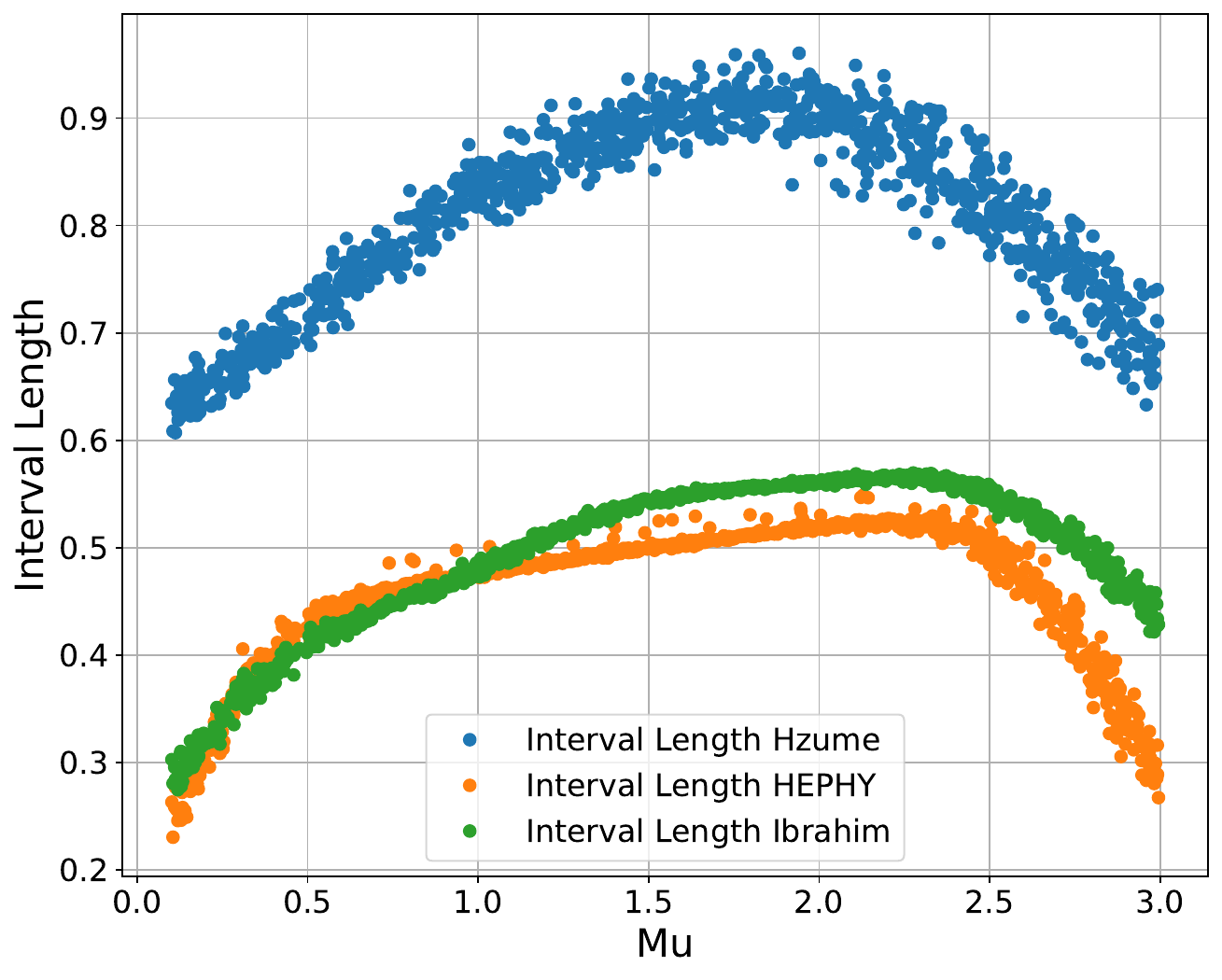}
        \caption{}
        \label{fig:interval_length_vs_mu}
    \end{subfigure}
    \hfill
    \begin{subfigure}[b]{0.32\textwidth}
        \includegraphics[width=\textwidth]{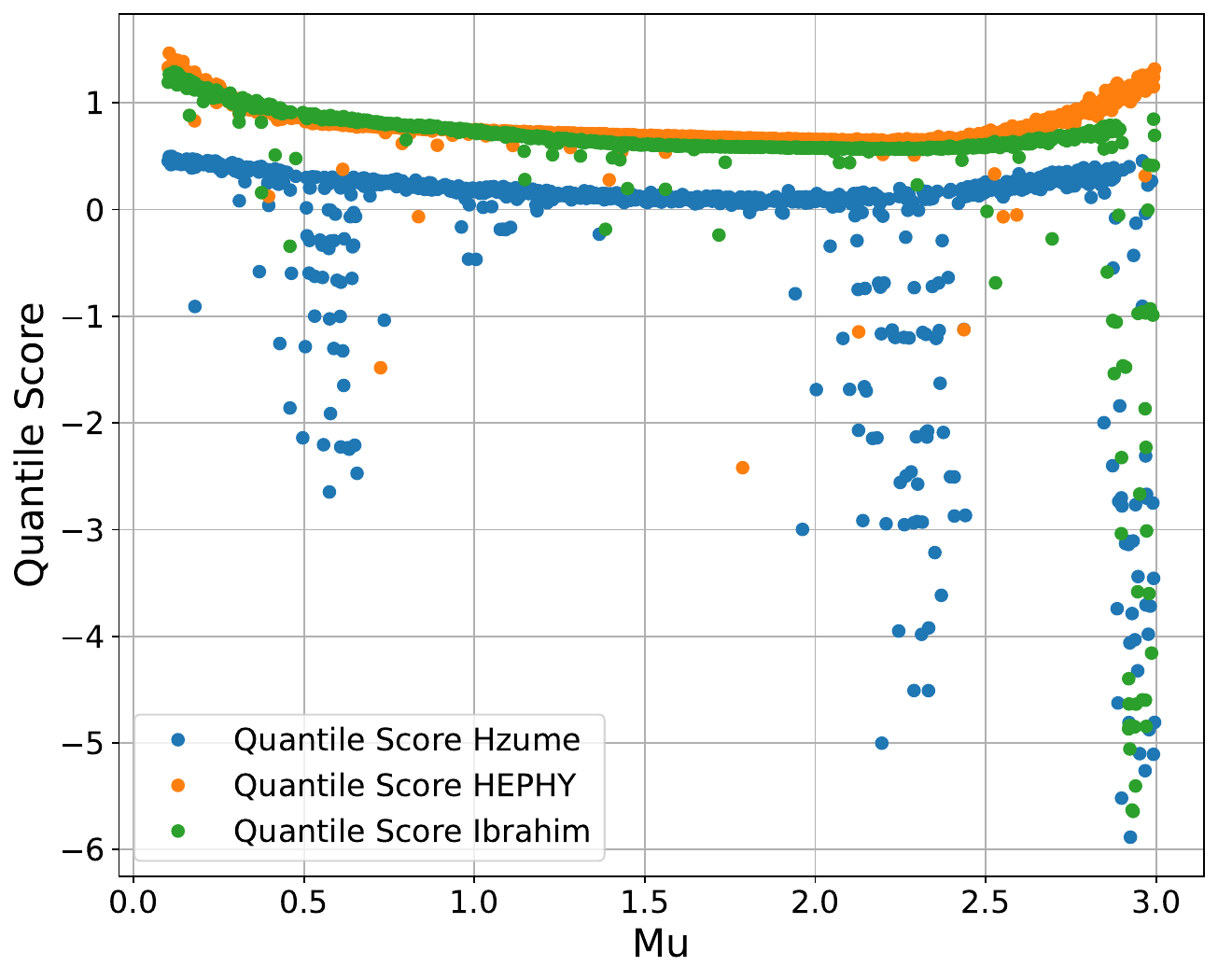}
        \caption{}
        \label{fig:quantile_score_vs_mu}
    \end{subfigure}
    \caption{Comparative study of the three finalists (blue for Hzume, orange for HEPHY and green for Ibrahim's model) with 1000 trials of 100 pseudo-experiments (see~\autoref{sec_metrics}). \ref{fig:coverage_vs_mu} the coverage from each trial, \ref{fig:interval_length_vs_mu} the average CI width  and \ref{fig:quantile_score_vs_mu} the quantile score }
    \label{fig:three_scores}
\end{figure}

\begin{itemize}
\item 1st tie: HEPHY (Lisa Benato, Cristina Giordano, Claudius Krause, Ang Li,  Robert Schöfbeck, Maryam Shooshtari, Dennis Schwarz, Daohan Wang) from Vienna’s Institute of High Energy Physics (HEPHY) in Austria wins \$2000.
\item 1st tie IBRAHIME (Ibrahim Elsharkawy) from University of Illinois at Urbana-Champaign, USA wins \$2000.
\item  3rd HZUME (Hashizume Yota) from Kyoto University, Japan wins \$500
\end{itemize}
All three are co-authors of this paper and have provided a summary of their algorithms in the following sub-sections. HEPHY and Ibrahim's sub-sections also refer to their public full papers and their code.

\input{hephy}
\input{IBRAHIME}
\input{HZUME}

\section{Conclusions and Outlook}
\label{sec:conclusion}

We have prepared a dataset~\cite{https://doi.org/10.5281/zenodo.15131565} (with relevant software~\cite{FAIR_Universe_dataset}), challenge, and platform for developing and comparing machine learning methods that quantify uncertainties in addition to providing point estimates.  
With the growing size of datasets in high-energy physics, the sophistication of tools, and the precision requirements to explore new phenomena, uncertainty quantification will be an essential part of machine learning in the future. The two winning approaches, \autoref{subsec:hephy}~\cite{Benato:2025rgo} and \autoref{subsec:ibrahime}~\cite{elsharkawy2025contrastivenormalizingflowsuncertaintyaware}, show two alternative techniques on how the treatment of systematic uncertainties can be incorporated successfully in experimental analyses.

The two techniques have very similar performances, however their results are not very correlated which implies the optimum has not been reached yet. Beyond this specific metric, we expect that this unique large dataset equipped with a biasing script will be the basis of future studies, for example: (i) the precise parametrisation of density and density ratios over several order of magnitudes which is fundamental to precision physics (ii) development of morphing/Optimal Transport techniques to parameterise multidimensional non-parametric biases (iii) the same studies but with a focus on learning with a limited number of instances. 




\section*{Acknowledgements}


We are grateful to the US Department of Energy, Office of High Energy Physics, and the subprogram on Computational High Energy Physics, for sponsoring this research, as well as to the ANR Chair of Artificial Intelligence HUMANIA (ANR-19-CHIA-0022). Seminal discussions contributing to this work took place at the workshop “Artificial Intelligence and the Uncertainty Challenge in Fundamental Physics,” sponsored by the CNRS AISSAI Center and the DATAIA Institute, and hosted at Institut Pascal at Université Paris-Saclay. The DATAIA Institute and Institut Pascal are respectively funded by the “Investissements d’Avenir” programs ANR-17-CONV-003 and ANR-11-IDEX-0003-01. This research used resources of the National Energy Research Scientific Computing Center (NERSC), a Department of Energy Office of Science User Facility using NERSC award HEP-ERCAP0032917. The computational results of \autoref{subsec:hephy}~\cite{Benato:2025rgo} were obtained using the CLIP cluster.

\david{cite whitepaper}
\bibliographystyle{tepml}
\bibliography{ref.bib,HEPML}
\newpage

\appendix

\section{Proton collisions and detection}
\label{sec_collisions}

This appendix gives details on how the data was generated.

The LHC collides bunches of protons every 25 nanoseconds within each of its four
experiments.
Two colliding protons produce a small firework
in which part of the kinetic energy of the protons is converted into new
particles. Most resulting particles are very unstable and decay quickly
into a cascade of lighter particles. The ATLAS detector measures
properties of these surviving particles (the so-called \emph{final state}): the
\emph{type} of the particle (electron, photon, muon, etc.), its \emph{energy},
and the 3D \emph{direction} of the particle. Based on these properties, the
decayed parent particle's properties can be inferred, and the inference chain
continues until the heaviest primary particles are reached.

An online trigger system discards most of the bunch collisions containing uninteresting events. The trigger is a three-stage cascade classifier which decreases the event rate from $40\,000\,000$ to about $400$ per second. The selected $400$ events are saved on disk, producing about one billion events and three petabytes of raw data per year.


The different types of
particles or pseudo-particles of interest for the challenge are electrons,
muons, hadronic tau, jets, and missing transverse energy. Electrons, muons, and
taus are the three leptons\footnote{For the list of elementary particles and
  their families, we refer the reader to
  \url{http://www.sciencemag.org/content/338/6114/1558.full}.}  from the
standard model.

Electrons and muons live long enough to reach the detector, so
their properties (energy and direction) can be measured directly. Conversely, Taus decay almost immediately after their creation into either an
electron and two neutrinos, a muon and two neutrinos, or a bunch of hadrons (charged
particles) and a neutrino. The bunch of hadrons can be identified as a pseudo-particle called the hadronic tau. Jets are pseudo particles rather than real
particles; they originate from a high-energy quark or gluon and appear in
the detector as a collimated energy deposit associated with charged tracks. The
primary information provided for the challenge is the measured momenta (see \autoref{sec_relativity} for a short introduction to
special relativity) of all the particles of the event.

\begin{figure}[h!]
\begin{center}
\includegraphics[width=0.5\textwidth]{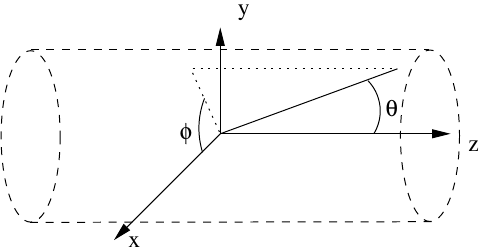}
\end{center}
\caption{ATLAS reference frame}
\label{fig_atlasframe}
\end{figure}

We are using the conventional 3D direct reference frame of ATLAS throughout the
document (see \autoref{fig_atlasframe}): the $z$ axis points along the horizontal
beam line, and the $x$ and $y$ axes are in the transverse plane with the $y$
axis pointing towards the top of the detector. $\theta$ is the polar angle and
$\phi$ is the azimuthal angle. Transverse quantities are quantities projected on
the $x - y$ plane, or, equivalently, quantities for which the $z$ component is
omitted. Instead of the polar angle $\theta$, we often use the {\it
  pseudorapidity} $\eta = -{\rm ln}\ {\rm tan} (\theta/2)$; $\eta = 0$
corresponds to a particle in the $x-y$ plane ($\theta = \pi/2$), $\eta =
+\infty$ corresponds to a particle traveling along the $z$-axis ($\theta=0$)
direction and $\eta = -\infty$ to the opposite direction ($\theta =
\pi$). Particles can be identified in the $\eta$  range in $[-2.5,2.5]$. For
$|\eta| \in [2.5,5]$, their momentum is still measured but they cannot be
identified. Particles with $|\eta|$ beyond~5 escape detection along the beam
pipe.

The missing transverse energy is a pseudo-particle which deserves a more
detailed explanation. The neutrinos produced in the decay of a tau escape
detection entirely. We can nevertheless infer their properties using the law
of momentum conservation by computing the vectorial sum of the momenta of all
the measured particles and subtracting it from the zero vector. In practice,
measurement errors for all particles make the sum poorly
estimated. Another difficulty is that many particles are lost in the beam pipe
along the $z$ axis, so the information on momentum balance is lost in the
direction of the $z$ axis. Thus, we can carry out the summation only in the
transverse plane, hence the name missing \emph{transverse} energy, which is a 2D
vector in the transverse plane.

\begin{figure}
\begin{center}
\includegraphics[width=0.6\textwidth]{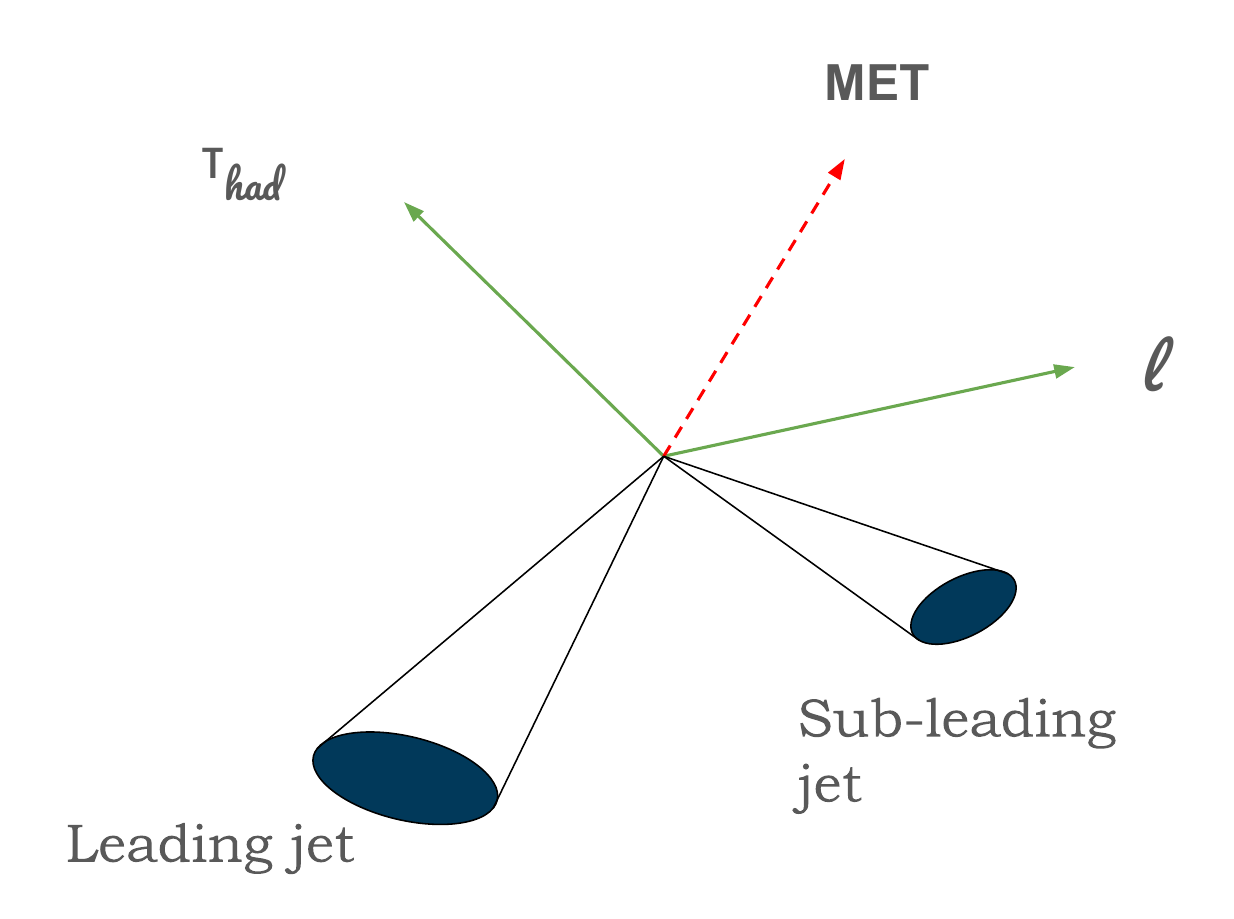}
\end{center}
\caption{Diagram of the particles in the final state chosen: one lepton, one tau hadron, up to two jets, and the missing transverse momentum vector, see text for details.}
\label{fig_particles_diag}
\end{figure}

For this competition, we selected only events with exactly one electron
or exactly one muon,  and with exactly one hadronic tau. These two particles should be 
of opposite electric charge. \autoref{fig_particles_diag} shows the particles in the selected final state, whose parameters are provided in the data.

To summarise, for each event, we produce a list of momenta for an electron or muon, a tau hadron, up to two jets, plus the missing transverse energy.

\begin{table}
    \centering
    \caption{Summary of the dataset for each category and subcategory. "Number Generated" is the number of events available in the dataset. In contrast, "LHC events" is the average number in this category in a pseudo-experiment corresponding to running of the Large Hadron Collider for 10 fb$^{-1}$, corresponding to approximately 800 billion inelastic proton collisions, or 2 weeks in summer 2024 conditions.}
        \renewcommand{\arraystretch}{1.5} 
        \begin{tabular}{rrrrr}
        \toprule
        Process & Number Generated & LHC Events & Label \\
            \midrule

            Higgs  & 52 040 227 & 1 015 & \textbf{\color{blue}signal}\\
            Z Boson & 160 383 358 & 1 002 395 & \textbf{\color{red}background}\\
            Di-Boson  & 605 118 & 3 783 &  \textbf{\color{red}background}\\
            $t \bar{t}$ & 7 070 398 & 44 192 & \textbf{\color{red}background}\\

            \bottomrule
        \end{tabular}
        \label{tab_data_table}
    \end{table}

Table~\ref{tab_data_table} details the number of events of each category in the dataset.

\newpage
\section{Special relativity}
\label{sec_relativity}

This appendix gives a very minimal introduction to special relativity for a
a better understanding of how the Higgs boson search is performed and what the
extracted features mean (taken mainly from \cite{higgsmldoc}).

\subsection{Momentum, mass, and energy}\label{sec_momentum}

A fundamental equation of special relativity defines the so-called 4-momentum of
a particle,
\begin{equation}\label{eq_emc2_1}
E^2=p^2c^2+m^2c^4,
\end{equation}
where $E$ is the energy of the particle, $p$ is its momentum, $m$ is the rest
mass and $c$ is the speed of light. When the particle is at rest, its momentum
is zero, and so Einstein's well-known equivalence between mass and energy,
$E=mc^2$, applies. In particle physics, we usually use the following units:
\unit{GeV} for energy, $\unit{GeV}/c$ for momentum, and $\unit{GeV}/c^2$ for
mass. \unit[1]{GeV} ($10^9$ electron-Volt) is one billion times the energy
acquired by an electron accelerated by a field of \unit[1]{V} over \unit[1]{m},
and it is also approximately the energy corresponding to the mass of a proton
(more precisely, the mass of the proton is about $\unit[1]{GeV}/c^2$). When
these units are used, \autoref{eq_emc2_1} simplifies to
\begin{equation}\label{eq_nemc2_2}
E^2=p^2+m^2.
\end{equation}
To avoid the clutter of writing $\unit{GeV}/c$ for momentum and $\unit{GeV}/c^2$
for mass, a shorthand of using \unit[GeV] for all the three quantities of
energy, momentum, and mass is usually adopted in most of the recent particle
physics literature (including papers published by the ATLAS and the CMS
experiments). We also adopt this convention throughout this document.

The momentum is related to the speed $v$ of the particle. For a particle with
non-zero mass, and when the speed of the particle is much smaller than the speed
of light $c$, the momentum boils down to the classical formula $p=mv$. In
special relativity, when the speed of the particle is comparable to $c$, we have
$p=\gamma m v$, where
\[
\gamma=\frac{1}{\sqrt{1-(v/c)^2}}.
\]
The relation holds both for the norms $v$ and $p$ and for the three dimensional
vectors $\vec{v}$ and $\vec{p}$, that is, $\vec{p}=\gamma m \vec{v}$, where, by
convention, $p = |\vec{p}|$ and $v = |\vec{v}|$. The factor $\gamma$ diverges to
infinity when $v$ is close to $c$, and the speed of light cannot be reached or
surpassed. Hence, momentum is a concept more frequently used than speed in
particle physics. The kinematics of a particle is fully defined by the momentum
and energy, more precisely, by the 4-momentum $(p_x,p_y,p_z,E)$. When a particle
is identified, it has a well-defined mass\footnote{neglecting the particle
  width}, so its energy can be computed from the
momentum and mass using \autoref{eq_emc2_1}. Conversely, the mass of a
particle with known momentum and energy can be obtained from
\begin{equation}\label{eq_inv_mass}
m=\sqrt{E^2-p^2}.
\end{equation}
Instead of specifying the momentum coordinate $(p_x,p_y,p_z)$, the parameters
$\phi$, $\eta$, and $p_\text{T}=\sqrt{p_x^2 + p_y^2}$, explained in
\autoref{sec_collisions} are often used.

\subsection{Invariant mass}

The mass of a particle is an intrinsic property of a particle. So, for all events
with a Higgs boson, the Higgs boson will have the same mass. To measure the mass
of the Higgs boson, we need the 4-momentum $(p_x,p_y,p_z,E) = (\vec{p},E)$ of
its decay products. Take the simple case of the Higgs boson $H$ decaying into a
final state of two particles, $A$ and $B$, which are measured in the detector. By
conservation of energy and momentum (which are fundamental laws of nature),
we can write $E_H=E_A+E_B$ and $\vec{p}_H=\vec{p}_A+\vec{p}_B$. Since the
energies and momenta of $A$ and $B$ are measured in the detector, we can compute
$E_H$ and $p_H = |\vec{p}_H|$ and calculate $m_H=\sqrt{E_H^2-p_H^2}$.  This is
called the {\em invariant mass} because (with a perfect detector) $m_H$ remains
the same even if $E_H$ and $p_H$ differ from event to event. This can be
generalised to more than two particles in the final state and to any number of
intermediate states.

In our case, the final state for particles originating from the Higgs boson is a lepton, a hadronic tau, and three
neutrinos. The lepton and hadronic tau are measured in the detector, but for the
neutrinos, all we have is the transverse missing energy, which estimates the sum of the momenta of the three neutrinos in the transverse plane. Hence, the mass of the $\tau\tau$ can
not be measured; we have to resort to different estimators which are only
correlated to the mass of the $\tau\tau$. For example, the {\em visible mass} (feature \texttt{\textbf{DER\_mass\_vis}})
which is the invariant mass of the lepton and the hadronic tau, hence
deliberately ignoring the unmeasured neutrinos. The possible jets in the events are not originating from the Higgs boson itself, but can be produced in association with it.


\subsection{Other useful formulas}

The following formulas are useful to compute DERived features from PRImary features
(in \autoref{sec_data_detailed}). For \texttt{tau}, \texttt{lep},
\texttt{leading\_jet}, and \texttt{subleading\_jet}, the momentum vector can be
computed as
\[
\vec{p}=\begin{pmatrix} p_x \\ p_y \\ p_z \end{pmatrix}=\begin{pmatrix}
p_\text{T} \times \cos \phi \\ p_\text{T} \times \sin \phi \\ p_\text{T} \times
\sinh \eta \end{pmatrix},
\]
where $p_\text{T}$ is the transverse momentum, $\phi$ is the azimuth angle,
$\eta$ is the pseudo rapidity, and $\sinh$ is the hyperbolic sine function. The
modulus of $p$ is
\begin{equation}\label{eq_mod}
p_\text{T} \times \cosh \eta,
\end{equation}
where $\cosh$ is the hyperbolic cosine function. The mass of these particles is
neglected, so $E = p$.

The missing transverse energy $\vec{E}_\text{T}^\text{miss}$ is a
two-dimensional vector
\[
\vec{E}_\text{T}^\text{miss} = \begin{pmatrix} |\vec{E}_\text{T}^\text{miss}|
  \times \cos \phi_\text{T} \\ |\vec{E}_\text{T}^\text{miss}| \times \sin
  \phi_\text{T} \end{pmatrix},
\]
where $\phi_\text{T}$ is the azimuth angle of the missing transverse
energy.

The invariant mass of two particles is the invariant mass of their 4-momentum sum,
that is (still neglecting the mass of the two particles),
\begin{equation}\label{eq_inv}
m_\text{inv}(\vec{a},\vec{b}) = \sqrt{\left(\sqrt{a_x^2+a_y^2+a_z^2} +
  \sqrt{b_x^2+b_y^2+b_z^2}\right)^2 - (a_x+b_x)^2 - (a_y+b_y)^2 - (a_z+b_z)^2.
}
\end{equation}
The transverse mass of two particles is the invariant mass of the vector sum, but this time
the third component is set to zero, which means only the projection on the transverse plane is considered.
That is (still neglecting the mass of the two particles),
\begin{equation}\label{eq_tr}
m_\text{tr}(\vec{a},\vec{b}) = \sqrt{\left(\sqrt{a_x^2+a_y^2} +
  \sqrt{b_x^2+b_y^2}\right)^2 - (a_x+b_x)^2 - (a_y+b_y)^2}.
\end{equation}
The pseudorapidity separation between two particles, $A$ and $B$, is
\begin{equation}\label{eq_eta_sep}
|\eta_A - \eta_B|.
\end{equation}
The $R$ separation between two particles $A$ and $B$ is
\begin{equation}\label{eq_r_sep}
\sqrt{(\eta_A - \eta_B)^2+(\phi_A - \phi_B)^2},
\end{equation}
where $\phi_A - \phi_B$ is brought back to the $]-\pi,+\pi]$ range. A good intuition for the $R$ separation is that it behaves like the 3D angle in radians between the two particles.

\clearpage
\section{The detailed description of the features}
\label{sec_data_detailed}

In this section, we explain the list of features that describe the events.

Prefix-less variables \texttt{Weight}, \texttt{Label},\texttt{DetailedLabel},
have a special role and should not be
used as regular features for the model\footnote{In the starting kit, they are split away in separate numpy arrays while the regular features are stored in a Dataframe}:

\begin{description}
\item[\texttt{\textbf{Weight}}] The event weight $w_i$. Not to be used as a feature. Not available in the test sample. 
\item[\texttt{\textbf{Label}}] The event label (integer) $y_i$ 1 for signal, 0 for background . Not to be used as a feature. Not available in the test sample. 
\item[\texttt{\textbf{DetailedLabel}}] The event detailed label (string) "htautau" for signal (when Label==1), "ztautau", "ttbar" and "diboson" for the three background categories (when Label==0). Not to be used as a feature. Not available in the test sample. This feature is used to implement some systematic biases; see \autoref{sec_syst}. It could be used to train a multi-category classifier.
\end{description}

The variables prefixed with \texttt{PRI} (for
PRImitives) are ``raw'' quantities about the bunch collision as measured by the
detector, essentially parameters of the momenta of particles (see \autoref{fig:PRI-1}, \autoref{fig:PRI-2} and \autoref{fig:PRI-3} for their distributions).   

In addition:

\begin{itemize}
\item Features are float unless specified otherwise.
\item All azimuthal $\phi$ angles are in radian in the $]-\pi,+\pi]$ range.
\item Energy, mass, and momentum are all in GeV
\item All other features are unitless.
\item Features are indicated as ``undefined'' when it can happen that
  they are meaningless or cannot be computed; in this case, their value is
  $-25$, which is outside the normal range of all variables.
\item The mass of particles has not been provided, as it can safely be neglected
  for the challenge.
\end{itemize}


\begin{description}
\item[\texttt{\textbf{PRI\_had\_pt}}] The transverse momentum $\sqrt{p_x^2 +
  p_y^2}$ of the hadronic tau.
\item[\texttt{\textbf{PRI\_had\_eta}}] The pseudorapidity $\eta$ of the hadronic
  tau.
\item[\texttt{\textbf{PRI\_had\_phi}}] The azimuth angle $\phi$ of the hadronic
  tau.
\item[\texttt{\textbf{PRI\_lep\_pt}}] The transverse momentum $\sqrt{p_x^2 +
  p_y^2}$ of the lepton (electron or muon).
\item[\texttt{\textbf{PRI\_lep\_eta}}] The pseudorapidity $\eta$ of the lepton.
\item[\texttt{\textbf{PRI\_lep\_phi}}] The azimuth angle $\phi$ of the lepton.
\item[\texttt{\textbf{PRI\_met}}] The missing transverse energy 
  $\vec{E}_\text{T}^\text{miss}$.
\item[\texttt{\textbf{PRI\_met\_phi}}] The azimuth angle $\phi$ of the missing
  transverse energy vector.
\item[\texttt{\textbf{PRI\_jet\_num}}] The number of jets.
\item[\texttt{\textbf{PRI\_jet\_leading\_pt}}] The transverse momentum
  $\sqrt{p_x^2 + p_y^2}$ of the leading jet, that is the jet with the largest
  transverse momentum (undefined if \texttt{PRI\_jet\_num}$\;=0$).
\item[\texttt{\textbf{PRI\_jet\_leading\_eta}}] The pseudorapidity $\eta$ of the
  leading jet (undefined if \texttt{PRI\_jet\_num}$\;=0$).
\item[\texttt{\textbf{PRI\_jet\_leading\_phi}}] The azimuth angle $\phi$ of the
  leading jet (undefined if \texttt{PRI\_jet\_num}$\;=0$).
\item[\texttt{\textbf{PRI\_jet\_subleading\_pt}}] The transverse momentum
  $\sqrt{p_x^2 + p_y^2}$ of the sub leading jet, that is, the jet with the second
  largest transverse momentum (undefined if \texttt{PRI\_jet\_num}$\;\leq1$).
\item[\texttt{\textbf{PRI\_jet\_subleading\_eta}}] The pseudorapidity $\eta$ of
  the subleading jet (undefined if \texttt{PRI\_jet\_num}$\;\leq\ 1$).
\item[\texttt{\textbf{PRI\_jet\_subleading\_phi}}] The azimuth angle $\phi$ of
  the subleading jet (undefined if \texttt{PRI\_jet\_num}$\;\leq\ 1$).
\item[\texttt{\textbf{PRI\_jet\_all\_pt}}] The scalar sum of the transverse
  momentum of all the jets of the events (not limited to the first 2).
\end{description}

\input{subplots_PRI}
\FloatBarrier

Variables prefixed with
\texttt{DER} (for DERived) are quantities computed from the primitive
features on the fly from PRImary features (including possible systematics shifts  )\footnote{The code to compute DERived features from PRImitive features can be seen at \url{https://github.com/FAIR-Universe/FAIR_Universe_dataset/blob/main/hep_challenge/derived_quantities.py}}(see \autoref{fig:DER-1} and \autoref{fig:DER-2} for their distributions). These quantities were selected by the physicists of ATLAS in the
reference document~\cite{Aad:2015vsa} either to select regions of interest or as features for the Boosted Decision Trees used in this analysis in order to enhance signal Higgs boson events separation from background events. DERived features were already present in the HiggsML dataset~\cite{higgsmldoc}\footnote{The notable exception of \texttt{DER\_mass\_MMC} which was in the HiggsML dataset but is deliberately absent from the Fair-Universe dataset because it was the result of a complex and lengthy Monte-Carlo Markov Chain integration which is not practical to rerun.}). The DERived features correspond to feature engineering; an ideal model to be trained on infinite statistics should not need these features. This distinction between primary and derived features (or "low-level" and "high-level" or "raw variables" and "human-assisted variables") is rather standard in the AI for HEP literature, see for example \cite{Baldi:2014kfa,BaldiTauTau}. There is no guarantee that all DERived features are useful for this challenge (they could even be detrimental in the context of systematics). The challenge participant is free to keep these DERived features, remove them altogether, keep a few, or do more feature engineering. 

\begin{description}

\item[\texttt{\textbf{DER\_mass\_transverse\_met\_lep}}] The transverse
  mass~(\autoref{eq_tr}) between the missing transverse energy and the lepton.
\item[\texttt{\textbf{DER\_mass\_vis}}] The invariant mass~(\autoref{eq_inv}) of the
  hadronic tau and the lepton.
\item[\texttt{\textbf{DER\_pt\_h}}] The modulus~(\autoref{eq_mod}) of the vector sum
  of the transverse momentum of the hadronic tau, the lepton, and the missing
  transverse energy vector.
\item[\texttt{\textbf{DER\_deltaeta\_jet\_jet}}] The absolute value of the pseudorapidity
  separation (\autoref{eq_eta_sep}) between the two jets (undefined if
  \texttt{PRI\_jet\_num}$\;\leq 1$).
\item[\texttt{\textbf{DER\_mass\_jet\_jet}}] The invariant mass~(\autoref{eq_inv})
  of the two jets (undefined if \texttt{PRI\_jet\_num}$\;\leq 1$).
\item[\texttt{\textbf{DER\_prodeta\_jet\_jet}}] The
  product of the pseudorapidities of the two jets (undefined if
  \texttt{PRI\_jet\_num}$\;\leq 1$).
\item[\texttt{\textbf{DER\_deltar\_had\_lep}}] The $R$
  separation~(\autoref{eq_r_sep}) between the hadronic tau and the lepton.
\item[\texttt{\textbf{DER\_pt\_tot}}] The modulus~(\autoref{eq_mod}) of the vector
  sum of the missing transverse momenta and the transverse momenta of the
  hadronic tau, the lepton, the leading jet (if \texttt{PRI\_jet\_num}$\;\geq
  1$) and the subleading jet (if \texttt{PRI\_jet\_num}$\;=2$) (but not of any
  additional jets).
\item[\texttt{\textbf{DER\_sum\_pt}}] The sum of the moduli~(\autoref{eq_mod}) of
  the transverse momenta of the hadronic tau, the lepton, the leading jet (if
  \texttt{PRI\_jet\_num}$\;\geq 1$) and the subleading jet (if
  \texttt{PRI\_jet\_num}$\;=2$) and the other jets (if
  \texttt{PRI\_jet\_num}$\;>=3$).
\item[\texttt{\textbf{DER\_pt\_ratio\_lep\_tau}}] The ratio of the transverse
  momenta of the lepton and the hadronic tau.
\item[\texttt{\textbf{DER\_met\_phi\_centrality}}] The centrality of the
  azimuthal angle of the missing transverse energy vector w.r.t.\ the hadronic
  tau and the lepton
  \[
  C = \frac{A+B}{\sqrt{A^2 + B^2}},
  \]
  where $A = \sin(\phi_\text{met} - \phi_\text{lep})*\text {sign}(\sin(\phi_\text{had}
  - \phi_\text{lep}))$, $B = \sin(\phi_\text{had} -
  \phi_\text{met})*\text {sign}(\sin(\phi_\text{had} - \phi_\text{lep}))$, and
  $\phi_{met}$, $\phi_\text{lep}$, and $\phi_\text{had}$ are the azimuthal
  angles of the missing transverse energy vector, the lepton, and the hadronic
  tau, respectively. The centrality is $\sqrt{2}$ if the missing transverse
  energy vector $\vec{E}_\text{T}^\text{miss}$ is on the bisector of the
  transverse momenta of the lepton and the hadronic tau. It decreases to 1 if
  $\vec{E}_\text{T}^\text{miss}$ is collinear with one of these vectors and it
  decreases further to $-\sqrt{2}$ when $\vec{E}_\text{T}^\text{miss}$ is
  exactly opposite to the bisector. The logic behind this feature is that if the neutrinos are colinear to the lepton and the hadronic tau (which is a good approximation), then the missing transverse energy vector should be between the lepton and the hadronic tau.
\item[\texttt{\textbf{DER\_lep\_eta\_centrality}}] The centrality of the
  pseudorapidity of the lepton w.r.t. the two jets (undefined if
  \texttt{PRI\_jet\_num}$\;\leq 1$)
  \[
  \exp \left[ \frac{-4}{(\eta_1-\eta_2)^2}\left( \eta_\text{lep} -
    \frac{\eta_1+\eta_2}{2} \right) ^2 \right],
  \]
  where $\eta_\text{lep}$ is the pseudorapidity of the lepton and $\eta_1$ and
  $\eta_2$ are the pseudorapidities of the two jets. The centrality is 1 when
  the lepton is on the bisector of the two jets, decreases to $1/e$ when it is
  collinear to one of the jets, and decreases further to zero at infinity. The logic behind this feature is that if the two jets are emitted together with the Higgs boson, then the Higgs decay product should be in average between the two jets.

 \end{description}

\input{subplots_DER}
 
The feature list and event sample are primarily inspired from \cite{Aad:2015vsa}. One crucial difference is that the dataset was produced with a more straightforward (leading-order) event generator (Pythia), and the detector effect was simulated with a more straightforward detector simulation (Delphes rather than Geant4 ATLAS Simulation). 
 These simplifications allowed us to provide to participants a large sample allowing the development of 
sophisticated models while preserving the complexity of the original
problem.

\clearpage

\section{Systematic biases}
\label{sec_syst}

This appendix details the implementation of the systematic biases Nuisance Parameters\footnote{See also \url{https://github.com/FAIR-Universe/FAIR_Universe_dataset/blob/main/hep_challenge/systematics.py}}. 

\subsection{Systematic bias definition}
\autoref{tab_sys} lists the different Nuisance Parameters with their Gaussian distribution and the range to which they are clipped. \ates, \ajes, and \asoftmet impacts some PRImary features, and then DERived features in cascade. \ates and \ajes also impact which events make it to the final dataset.
\attbar, \adiboson and \abkg only impact the Weight of some background categories, that is to say, the composition of the background (for \attbar and \adiboson) or the overall level of the background \abkg. The Gaussian distributions parameterise our ignorance of the exact value of the biases. We think their value is 1 (or zero for \asoftmet) while their real value is slightly different, as parameterised by their width, thus biasing our measurement by an unknown amount, which can be simulated.

\begin{table}
\begin{center}
\begin{tabular}{|c|c|c|c|}
\hline
Variable & Mean & Sigma & Range \\
\hline
\ates  & 1. & 0.01 & [0.9, 1.1] \\
\ajes & 1. & 0.01 & [0.9, 1.1] \\
\asoftmet & 0. & 1. & [0., 5.] \\
\attbar & 1. & 0.02 & [0.8, 1.2] \\
\adiboson & 1.& 0.25 & [0., 2.] \\
\abkg & 1. & 0.001 & [0.99, +1.01] \\
\hline
\end{tabular}
\end{center}
\caption{\label{tab_sys} List of six systematic bias Nuisance Parameters defined in the challenge, with the mean and sigma of their Gaussian (Log-normal for \asoftmet) distribution and their range. The corresponding $\alpha$ is set to the Mean value whenever a systematic bias is switched off. "No systematics" means all $\alpha$ are set to their Mean value.}
\end{table}

\subsection{Impact of biases on features}



To detail the impact of the systematics, we need to detail first how the 4-momenta from the final state particles can be reconstructed from the PRImary features, following \autoref{sec_relativity}. The four parameters ($P_x$,$P_y$,$P_z$,E) of the four-vector of each particle in the final state can be reconstructed from the PRImary features as follows (using the hadronic tau as an example, and reminding that the mass is neglected so that $E=P$),
\newcommand{\phad}{\ensuremath{P_{\rm had}}\xspace}
\newcommand{\plep}{\ensuremath{P_{\rm lep}}\xspace}
\newcommand{\pljet}{\ensuremath{P_{\rm leading\ jet}}\xspace}
\newcommand{\psljet}{\ensuremath{P_{\rm subleading\ jet}}\xspace}
\newcommand{\pmet}{\ensuremath{P_{\rm MET}}\xspace}

\[
\phad = \left( \begin{array}{c}
\texttt{PRI\_had\_pt} * \cos(\texttt{PRI\_had\_phi}) \\
\texttt{PRI\_had\_pt} * \sin(\texttt{PRI\_had\_phi}) \\
\texttt{PRI\_had\_pt} * \sinh(\texttt{PRI\_had\_eta})\\
\texttt{PRI\_had\_pt} * \cosh(\texttt{PRI\_had\_eta}) \\
\end{array} \right)
\]
(where $\sinh$ and $\cosh$ are the hyperbolic sine and cosine functions), and similarly for \plep, \pljet and \psljet.

The Missing ET vector is, by definition, in the transverse plane, so we have: 
\[
\pmet = \left( \begin{array}{c}
\texttt{PRI\_met} * \cos(\texttt{PRI\_met\_phi}) \\
\texttt{PRI\_met} * \sin(\texttt{PRI\_met\_phi})\\
\texttt{PRI\_met} \\
\end{array} \right)
\]

\ates is meant to describe the fact that the detector is not calibrated correctly for the measurement of the hadron momentum, meaning when the detector reports a momentum $P_{\rm had}$ it really is : 
\[
P_{\rm had}^{\rm biased}= \ates P_{\rm had} 
\]
And similarly, for the jets momentum (when they are defined)
\[
P_{\rm jet\_leading}^{\rm biased}= \ajes P_{\rm jet\_leading}
\]
\[
P_{\rm jet\_subleading}^{\rm biased}= \ajes P_{\rm jet\_subleading}
\]
\ates and \ajes also have an impact on \pmet : \pmet is obtained from the opposite of the sum of all visible objects in the event so that changing one of the visible objects (like \phad, \pljet or \psljet) has a correlated impact on \pmet (this calculation is performed on the first two coordinates and $E_{\rm MET}$ is recalculated from their modulus):

\[
\pmet^{biased}=\pmet+(1-\ates)\phad+(1-\ajes)\pljet+(1-\ajes)\psljet 
\]

\asoftmet has a different role; it expresses an additional noise source in the measurement of the missing ET vector, which is not present in the simulation. A random 2D vector of norm $ET_{\rm soft}={\rm Lognormal(\asoftmet)}$ is added to \pmet  (with different values event by event, by contrast with \asoftmet, which has a fixed value for a given pseudo-experiment) (this calculation is performed on the first two coordinates and $E_{\rm MET}$ is recalculated from their modulus):
\[
\pmet^{biased}=\pmet+ \left( \begin{array}{c}
Gauss (0,ET_{\rm soft}) \\
Gauss (0,ET_{\rm soft}) \\
\end{array} \right)
\]

The corresponding modified PRImary features are then recomputed to new biased values: PRI\_had\_pt, PRI\_leading\_jet\_pt, PRI\_leading\_jet\_pt, PRI\_met, and PRI\_met\_phi.  

In addition, 
\[
PRI\_jet\_all\_pt^{\rm biased} = \ajes \times PRI\_jet\_all\_pt
\]
If the number of jets is three or more, the impact of \ajes on missing ET cannot be calculated, given that detailed information on the additional jets (beyond two) is not available; this is a legitimate approximation as the total jet transverse momentum would be in most cases dominated by the first two leading.  

DERived features are also impacted if they depend on these PRImary features (see \autoref{sec_data_detailed}). Thus, for each of \ates, \ajes and \asoftmet, different features are impacted in a correlated way.

\subsection {Weight impacting bias implementation}

\abkg, \attbar and \adiboson only impact the Weight of background events, more precisely:
\begin{itemize}
\item events with \texttt{DetailedLabel="ztautau"}:
\[
{\rm Weight}^{bias}=\abkg \times {\rm Weight}
\]
\item events with \texttt{DetailedLabel="ttbar"}:
\[
{\rm Weight}^{bias}=\abkg \times \attbar \times {\rm Weight}
\]
\item events with \texttt{DetailedLabel="diboson"}:
\[
{\rm Weight}^{bias}=\abkg \times \adiboson \times  {\rm Weight}
\]
\end{itemize}   

So \abkg only affects the overall level of the background but leaves the background distributions unchanged. \attbar and \adiboson impacts only the proportion of the smaller backgrounds (see \autoref{tab_data_table}), thus distorting the overall background distribution.

\subsection{Event selection}

Hadronic tau (and also the jets) can only be identified in the detector above a certain transverse momentum threshold ("low threshold" in the following) so that the raw dataset PRI\_had\_pt, PRI\_jet\_leading\_pt PRI\_jet\_subleading\_pt  have clear thresholds. When applying \ates and \ajes, these thresholds move so that if nothing else is done, the threshold position would be an obvious giveaway of the value of \ates and \ajes. 

To alleviate this, "high thresholds" (see \autoref{tab_threshold}) have been defined, which should systematically be applied after the calculation of the biased PRImary parameters, so that the thresholds to be observed on PRI\_had\_pt, PRI\_jet\_leading\_pt PRI\_jet\_subleading\_pt are independent of \ates and \ajes. The ranges in \autoref{tab_sys} are such that the thresholds should also be applied when no systematics bias is used\footnote{In practice, function \texttt{systematics} in \url{https://github.com/FAIR-Universe/FAIR_Universe_dataset/blob/main/hep_challenge/systematics.py} should always be used, even in the no systematics case.}.

\begin{table}
\begin{center}
\begin{tabular}{|c|c|c|}
\hline
Variable & Low threshold (GeV) & High threshold (GeV) \\
\hline
$\phad^{\rm T}$  & $\simeq$ 23 & 26 \\
$\pljet^{\rm T}$ and $\psljet^{\rm T}$ & $\simeq$ 23 & 26 \\
\hline
\end{tabular}
\end{center}
\caption{\label{tab_threshold} Low and high threshold of hadronic tau and jet transverse momentum.  }
\end{table}

\end{document}
\newpage
\section*{NeurIPS Paper Checklist}

\begin{enumerate}

\item {\bf Claims}
    \item[] Question: Do the main claims made in the abstract and introduction accurately reflect the paper's contributions and scope?
    \item[] Answer: \answerYes{} 
    \item[] Justification: 

\item {\bf Limitations}
    \item[] Question: Does the paper discuss the limitations of the work performed by the authors?
    \item[] Answer: \answerYes{} 
    \item[] Justification: see \autoref{sec_limitations}

\item {\bf Theory assumptions and proofs}
    \item[] Question: For each theoretical result, does the paper provide the full set of assumptions and a complete (and correct) proof?
    \item[] Answer: \answerNA{} 
    \item[] Justification: no theoretical result 

    \item {\bf Experimental result reproducibility}
    \item[] Question: Does the paper fully disclose all the information needed to reproduce the main experimental results of the paper to the extent that it affects the main claims and/or conclusions of the paper (regardless of whether the code and data are provided or not)?
    \item[] Answer: \answerNo{} 
    \item[] Justification: given the page allocation, only short summaries of the methods of the winning trios could be provided. Code of the two winners is provided though. 

\item {\bf Open access to data and code}
    \item[] Question: Does the paper provide open access to the data and code, with sufficient instructions to faithfully reproduce the main experimental results, as described in supplemental material?
    \item[] Answer: \answerYes{} 
    \item[] Justification: see dataset~\cite{https://doi.org/10.5281/zenodo.15131565} and  software~\cite{FAIR_Universe_dataset}. Code to reproduce the dataset is also available \cite{FAIR_Universe_dataGen}.  

\item {\bf Experimental setting/details}
    \item[] Question: Does the paper specify all the training and test details (e.g., data splits, hyperparameters, how they were chosen, type of optimizer, etc.) necessary to understand the results?
    \item[] Answer: \answerNo{}{} 
    \item[] Justification: given the page allocation, only short summaries of the methods of the winning trios could be provided. Code of the two winners is provided though. 

\item {\bf Experiment statistical significance}
    \item[] Question: Does the paper report error bars suitably and correctly defined or other appropriate information about the statistical significance of the experiments?
    \item[] Answer: \answerYes{}{} 
    \item[] Justification: In \autoref{sec:results}, we report the final score of the winning trio HEPHY -0.582, Ibrahim -0.576 and
HZUME -2.16. There is no meaningful error bar to quote on these numbers because, given that they are measured on the same pseudo-experiment, they are very correlated. An additional bootstrap analysis (which could not be detailed given page allocation) showed that HEPHY
and Ibrahim could not be reliably ranked, hence we had to declare a tie. Hence the Yes to this question. 

\item {\bf Experiments compute resources}
    \item[] Question: For each experiment, does the paper provide sufficient information on the computer resources (type of compute workers, memory, time of execution) needed to reproduce the experiments?
    \item[] Answer: \answerYes{}{} 
    \item[] Justification: the compute resources to simulate the dataset, and to train the trio's models are reported (please search for "hour"). Model inference was limited to 20s per pseudo-experiment. 
    
\item {\bf Code of ethics}
    \item[] Question: Does the research conducted in the paper conform, in every respect, with the NeurIPS Code of Ethics \url{https://neurips.cc/public/EthicsGuidelines}?
    \item[] Answer: \answerYes{}
    \item[] Justification: 

\item {\bf Broader impacts}
    \item[] Question: Does the paper discuss both potential positive societal impacts and negative societal impacts of the work performed?
    \item[] Answer: \answerYes{}{} 
    \item[] Justification: see Conclusion~\autoref{sec:conclusion} 
    
\item {\bf Safeguards}
    \item[] Question: Does the paper describe safeguards that have been put in place for responsible release of data or models that have a high risk for misuse (e.g., pretrained language models, image generators, or scraped datasets)?
    \item[] Answer: \answerNo{}{} 
    \item[] Justification: we could not think of possible mis-use of our dataset

\item {\bf Licenses for existing assets}
    \item[] Question: Are the creators or original owners of assets (e.g., code, data, models), used in the paper, properly credited and are the license and terms of use explicitly mentioned and properly respected?
    \item[] Answer: \answerYes{} 
    \item[] Justification: all existing assets have been cited according to common practice 

\item {\bf New assets}
    \item[] Question: Are new assets introduced in the paper well documented and is the documentation provided alongside the assets?
    \item[] Answer: \answerYes{} 
    \item[] Justification: see \url{https://zenodo.org/records/15131565}. This paper will also be added as supplementary documentation to the zenodo record.

\item {\bf Crowdsourcing and research with human subjects}
    \item[] Question: For crowdsourcing experiments and research with human subjects, does the paper include the full text of instructions given to participants and screenshots, if applicable, as well as details about compensation (if any)? 
    \item[] Answer: \answerNA{} 
    \item[] Justification: we assume a scientific competition does not count as crowdsourcing 

\item {\bf Institutional review board (IRB) approvals or equivalent for research with human subjects}
    \item[] Question: Does the paper describe potential risks incurred by study participants, whether such risks were disclosed to the subjects, and whether Institutional Review Board (IRB) approvals (or an equivalent approval/review based on the requirements of your country or institution) were obtained?
    \item[] Answer: \answerNA{} 
    \item[] Justification: 

\item {\bf Declaration of LLM usage}
    \item[] Question: Does the paper describe the usage of LLMs if it is an important, original, or non-standard component of the core methods in this research? Note that if the LLM is used only for writing, editing, or formatting purposes and does not impact the core methodology, scientific rigorousness, or originality of the research, declaration is not required.
    \item[] Answer: \answerNA{} 
    \item[] Justification: 

\end{enumerate}

\end{document}

%% file: hephy.tex
\subsection{\texttt{HEPHY}: Simulation-based inference with a calibrated multiclassifier and parametric regressors for learning systematics}
\label{subsec:hephy}
\providecommand{\PH}{\ensuremath{\mathrm{H}}\xspace} 
\providecommand{\PZ}{\ensuremath{\mathrm{Z}}\xspace} 
\newcommand{\ttbar}{\ensuremath{\mathrm{t}\overline{\mathrm{t}}}\xspace} 
\newcommand{\VV}{\ensuremath{\mathrm{VV}}\xspace}
\newcommand{\bx}{\ensuremath{\boldsymbol{x}}\xspace}
\newcommand{\ddd}{\ensuremath{{\textrm{d}}}\xspace}
\newcommand{\bn}{\ensuremath{{\boldsymbol{\nu}}}\xspace}
\newcommand{\bzero}{{\ensuremath{\mathbf{0}}}\xspace}

We use simulation-based inference (SBI) to construct a flexible, unbinned, and refinable likelihood model~\cite{Schofbeck:2024zjo} that captures the full high-dimensional event information for inference of the signal strength $\mu$ and the nuisance parameters \bn via a multiclassifier and parametric regressors~\cite{Benato:2025rgo}. The code-base for ``Guaranteed Optimal Likelihood-based Unbinned Method'' (\textsc{GOLLUM}) is publicly available at Ref.~\cite{GOLLUM}. Only a brief description is provided here. In the extended profiled likelihood-ratio test-statistic 
$q_{\mu}(\mathcal{D})=-2\log\frac{\max_{\bn\phantom{,\bn}}L(\mathcal{D}|\mu,\bn)}{\max_{\mu,\bn}L(\mathcal{D}|\mu,\bn)}$, we introduce a reference likelihood to the nominal (unvaried) hypothesis as $q_{\mu}(\mathcal{D}) = \min_{\bn} u(\mathcal{D}|\mu,\bn) - \min_{\mu,\bn} u(\mathcal{D}|\mu,\bn)$ where $-\frac{1}{2}u(\mathcal{D}|\mu,\bn) = -\mathcal{L}(\sigma(\mu,\bn) - \sigma(1,\bzero))+ \sum_{i=1}^{N_\text{obs}}\log\left(\frac{\ddd\sigma(\bx_i|\mu,\bn)}{\ddd\sigma(\bx_i|1,\bzero)}\right)$. We parametrise the inclusive yield ($\mathcal{L}\sigma (\mu,\bn)$) (total number of expected events) and differential cross section ratio $\frac{\ddd\sigma(\bx|\mu,\bn)}{\ddd\sigma(\bx|1,\bzero)}$ (density ratios) by surrogates in six disjoint selections, two of which are signal-enriched and the rest serve to constrain nuisance parameters. 

A multiclass classifier is trained on nominal (i.e., unvaried) simulation data and predicts the class probabilities for the four processes: $\PH \to \tau\tau$, $\PZ \to \tau\tau$, $\ttbar$, and $\VV$. The output class probability is scaled with $(1+\alpha)^\nu$ for each of the nuisance parameters $\nu_{\text{bkg}}$, $\nu_{\ttbar}$, and $\nu_{\VV}$, that control the normalization of background processes. The three accompanying constants $\alpha$ determine the pre-fit sizes of these uncertainties.  A critical step is a dedicated and highly precise iterative isotonic regression step to calibrate the classifier's output. 

To account for the dependence of the likelihood on the remaining systematic uncertainties, a second set of networks estimates the relative variation in the differential cross section as a function of $\bn_\text{calib}=\{\nu_\text{tes}, \nu_\text{jes},  \nu_\text{met}\}$. These nuisances control uncertainties in the calibration of the data and enter training data via the biasing script. We fit an exponential ansatz parameterised by a neural network for each of the four processes (labeled by $p$) and separately in each region: $\frac{\ddd\sigma_p(\bx|\mu,\bn)}{\ddd\sigma_p(\bx|1,\bzero)}\simeq \hat S_p(\bx|\bn_\text{calib})=\exp(\nu_A \hat\Delta_{r,p,A}(\bx))$, where $\nu_A$ is a multi-index that labels three linear, three quadratic, and three mixed terms of the three calibration-type nuisances. The $\hat\Delta_{r,p,A}(\bx)$ are functions learned by the network and specific to the selection $r$ and the process $p$. Based on the cross-entropy loss, the ansatz leads to the loss function
\begin{align}
&L[\hat \Delta_A]=\sum_{\bn\in\mathcal{V}}\Bigg[\int\ddd\sigma(\bx|\bzero)\,\textrm{Soft}^+(\nu_A\hat\Delta_A(\bx))+\int\ddd\sigma(\bx|\bn)\,\textrm{Soft}^+(-\nu_A\hat\Delta_A(\bx))\Bigg].\label{eq:loss-analytic-SoftMax}
\end{align}
This architecture allows the surrogate to interpolate continuously in both feature and nuisance parameter space. The complete likelihood can then be computed from the surrogate for the differential cross-section ratio with the closed-form expression
\begin{align}
\frac{\ddd\sigma(\bx|\mu,\bn)}{\ddd\sigma(\bx|1,\bzero)\,}&\simeq\mu\hat g_\PH(\bx)\hat S_\PH(\bx|\bn_\text{calib})+(1+\alpha_\text{bkg})^{\nu_\text{bkg}}\Big(\hat g_\PZ(\bx)\hat S_\PZ(\bx|\bn_\text{calib})\nonumber\\
&+(1+\alpha_{\ttbar})^{\nu_{\ttbar}}\,\hat g_{\ttbar}(\bx)\hat S_{\ttbar}(\bx|\bn_\text{calib})+(1+\alpha_\VV)^{\nu_\VV}\,\hat g_{\VV}(\bx)\hat S_{\VV}(\bx|\bn_\text{calib})\Big)\label{eq:xsec-ratio-ML}
\end{align}
where $\hat g_p(\bx)$ is the (calibrated) output of the multiclassifier. The surrogate is efficient in evaluating and differentiable with respect to all parameters. For the inclusive cross-section component of the extended likelihood, we introduce a spline-based interpolation scheme that reduces numerical instabilities and speeds up the evaluation during profiling.

We train one multiclass classifier and one systematic network per selection. Closure tests show that the surrogates reproduce the shapes and normalisations of the simulated distributions across many kinematic observables and several orders of magnitude. The unbinned surrogate could be further refined in a modular way: new systematics or background processes can be added without retraining the entire model, mirroring the workflow of traditional HEP data analyses. This ``refinable'' modelling is crucial for scalability in real LHC analyses where hundreds of nuisance parameters are typical. 

\begin{figure}
    \centering
    \includegraphics[width=0.305\linewidth]{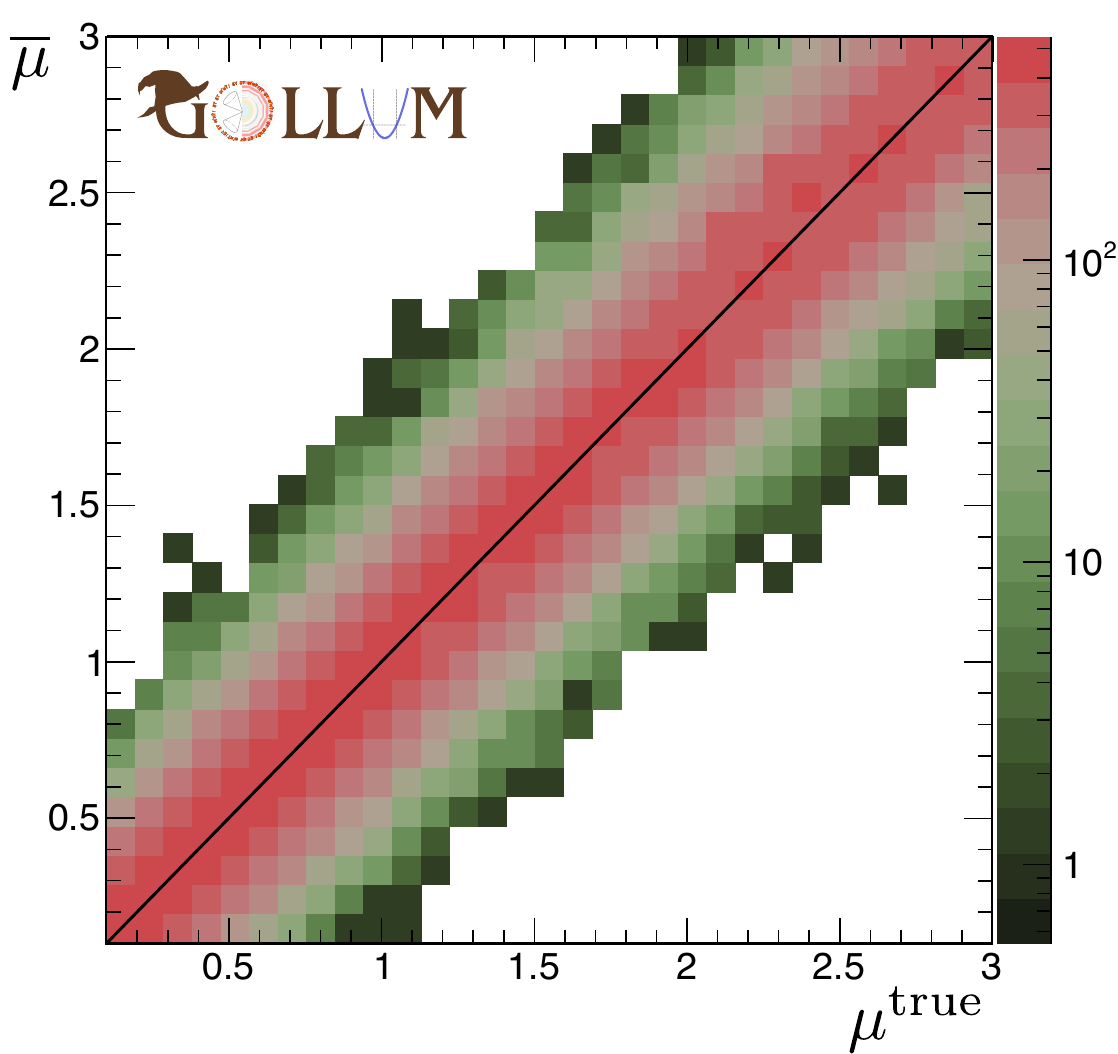}
    \includegraphics[width=0.32\linewidth]{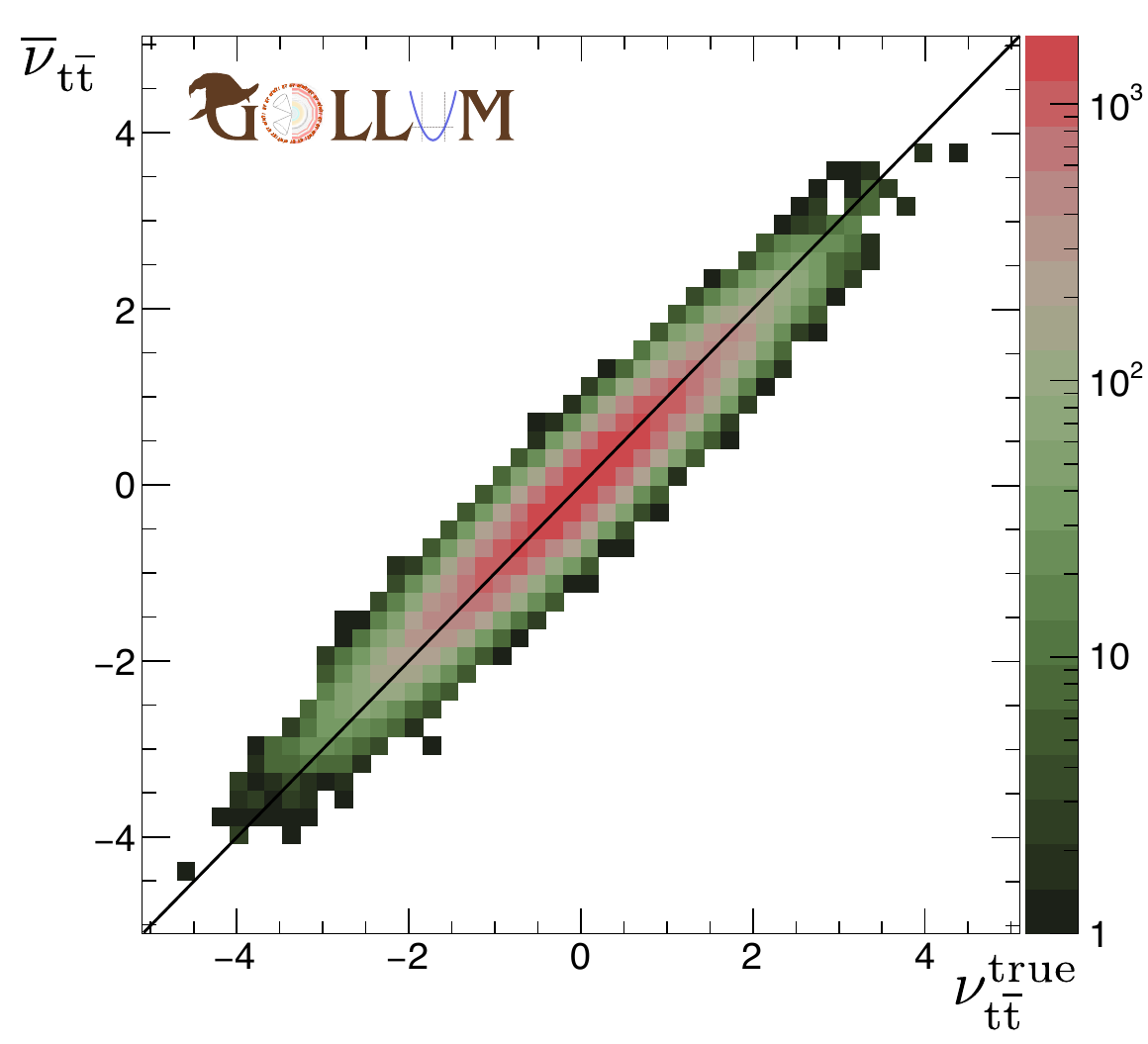}
    \includegraphics[width=0.32\linewidth]{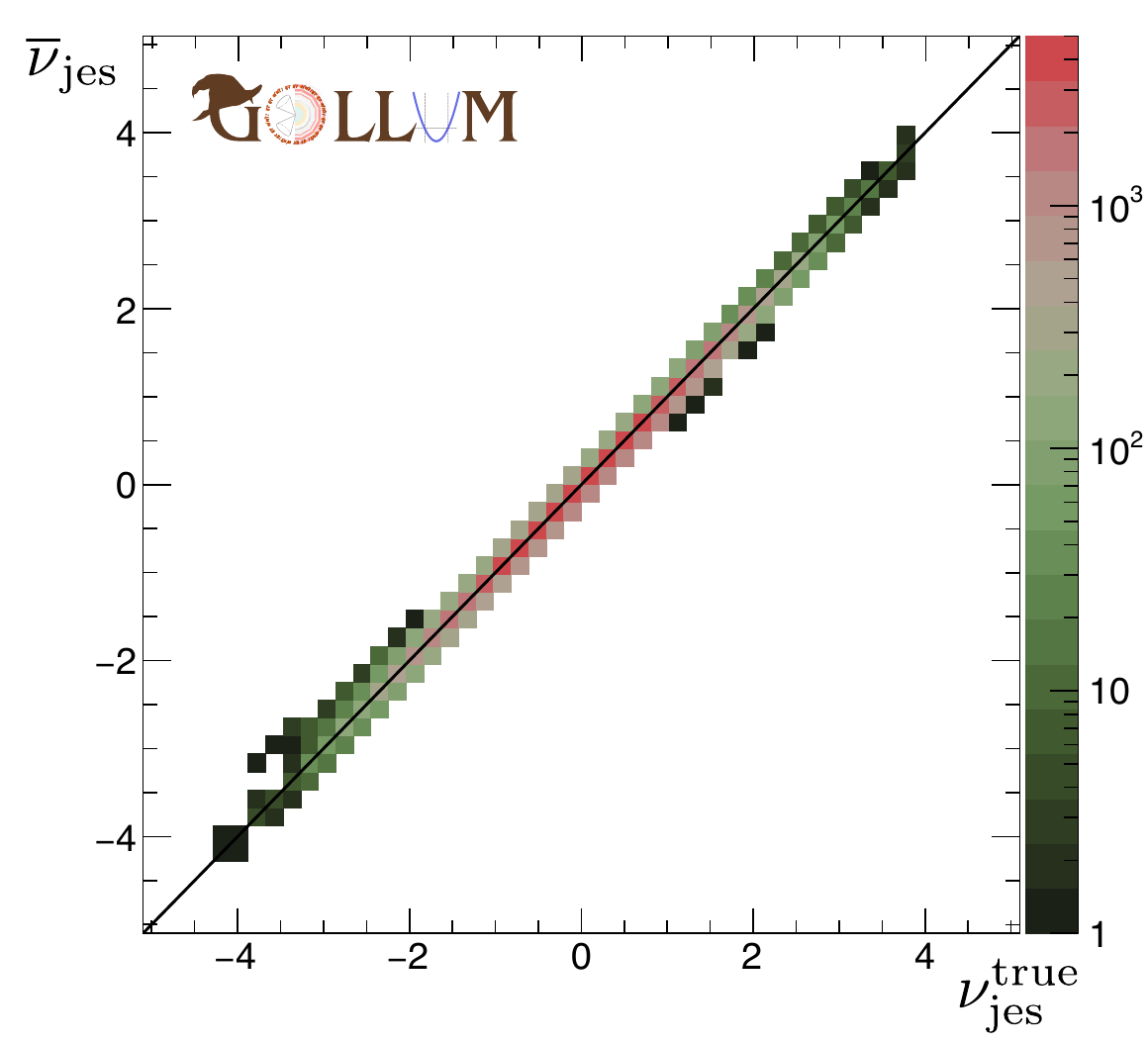}
    \caption{\label{fig:toy-scatter-mu} Scatter plot of the true value of the $\PH\to\tau\tau$ signal strength parameter $\mu$~(left) and the MLE $\bar\mu$ for $5\cdot10^4$ toys showing stability over the whole range of relevant $\mu_\text{true}$. The normalisation-type nuisance parameter $\nu_\text {\ttbar}$~(middle) and the calibration-type nuisance parameter $\nu_\text{jes}$~(right) are severely constrained, reducing the impact of the corresponding uncertainties. }\label{fig:corr}
\end{figure}

We profile the nuisance parameters using the \texttt{MINUIT} package~\cite{James:1975dr} and determine the 68\% CI by evaluating the profiled likelihood as a function of $\mu$. 
The gain from the unbinned model becomes evident when performing inference, where the unbinned surrogate model improves the expected $1\sigma$ confidence interval on the signal strength by 20\% compared to a traditional binned analysis using classifier-based templates. The unbinned model also leads to significantly stronger constraints on nuisance parameters, especially for calibration-related systematics like $\nu_{\text{tes}}$ and $\nu_{\text{jes}}$, reducing their impact on $\mu$ by up to 65\% when compared with the binned case~\cite{Benato:2025rgo}.

We assess the performance with $5\cdot10^4$ toys 
in~\autoref{fig:corr}. The signal strength $\mu$ is reconstructed stably over the whole range of relevant $\mu_\text{true}$. We severely constrain $\nu_\text {\ttbar}$ and $\nu_\text{jes}$, reducing the impact of the corresponding uncertainties. 
The total training time for the model was 200 CPU core hours.

%% file: ibrahime.tex
\subsection{\texttt{ibrahime}: Contrastive Normalizing Flows for Uncertainty-Aware Parameter Estimation}
\label{subsec:ibrahime}
The full description of the method can be found in the method paper \cite{elsharkawy2025contrastivenormalizingflowsuncertaintyaware}. The code used to train and evaluate the method is available at \cite{Elsharkawy2025CNFParameterEstimation}. 
\paragraph{Motivation}
A binary classifier can, in principle, estimate any model parameter $\Theta_i$ by learning a monotonic approximation of the likelihood ratio $r(\mathbf{x},\{\Theta_i,\nu_i\},\{\Theta'_i,\nu'_i\})
\propto \frac{P(\mathbf{x}\!\mid\!\{\Theta_i,\nu_i\})}{
P(\mathbf{x}\!\mid\!\{\Theta'_i,\nu'_i\})}$~\cite{Cranmer:2015bka}, where $\mathbf{x}$ are the data features and $\nu_i$ are nuisance parameters. In practice, this classifier approach can be impractical; if the number of model parameters $k_\Theta$ or nuisance parameters $k_\nu$ is large, the dimensionality prevents sufficient sampling of parameter space for many choices of $\{\Theta_i,\nu_i\}$. For the challenge, $\Theta\equiv\mu\propto f_s$, where $f_s$ is the signal fraction, and $\nu_i$ are the six HiggsML nuisance parameters. Given $\mu\propto f_s$ we can attempt to learn instead the likelihood ratio $r(\mathbf{x}, \{\nu_i\}, \{\nu'_i\})\propto\frac{p_s(\mathbf{x}|\{\nu_i\})}{p_{b}(\mathbf{x}|\{\nu'_i\})}$, where $p_s$ and $p_b$ are the signal and background distributions, by training on class labels and then determining $\mu$ with maximum likelihood estimation. 
To remedy the curse of dimensionality, we then replace the raw nuisance parameters $\nu_i$ with some \emph{discrimination functions} $\Phi_{s,b}[\mathbf{x};\{\nu_i\}]$ such that $r(\mathbf{x}, \{\nu_i\}, \{\nu'_i\})\propto\frac{p_s(\mathbf{x}|\Phi_s[\mathbf{x};\{\nu_i\}])}{p_{b}(\mathbf{x}|\Phi_b[\mathbf{x};\{\nu'_i\}])}.$ If these discrimination functions are relatively insensitive to nuisance parameters and take very different values for $\mathbf{x} \sim p_s$ compared to $\mathbf{x} \sim p_b$, a classifier trained on these features will more accurately approximate the desired likelihood with less data. We argue that Contrastive Normalising Flows (CNFs) are especially suitable for these functions $\Phi_{s,b}[\mathbf{x};\{\nu_i\}]$.  

\begin{figure}[t]
  \centering
  \begin{subfigure}[t]{0.39\linewidth}
      \centering
    \includegraphics[width=\linewidth]{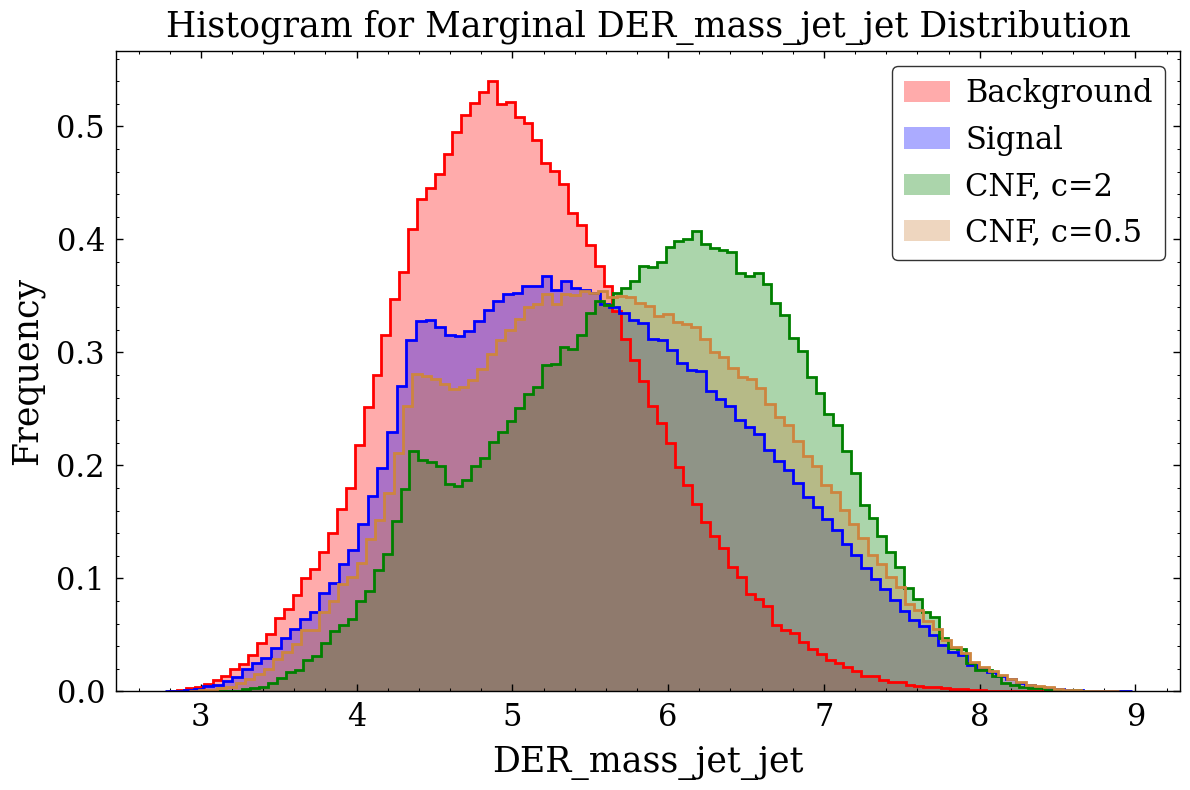}
    \label{fig:marginal_varyc}
    \end{subfigure}
\begin{subfigure}[t]{0.30\linewidth}
    \centering
    \includegraphics[width=\linewidth]{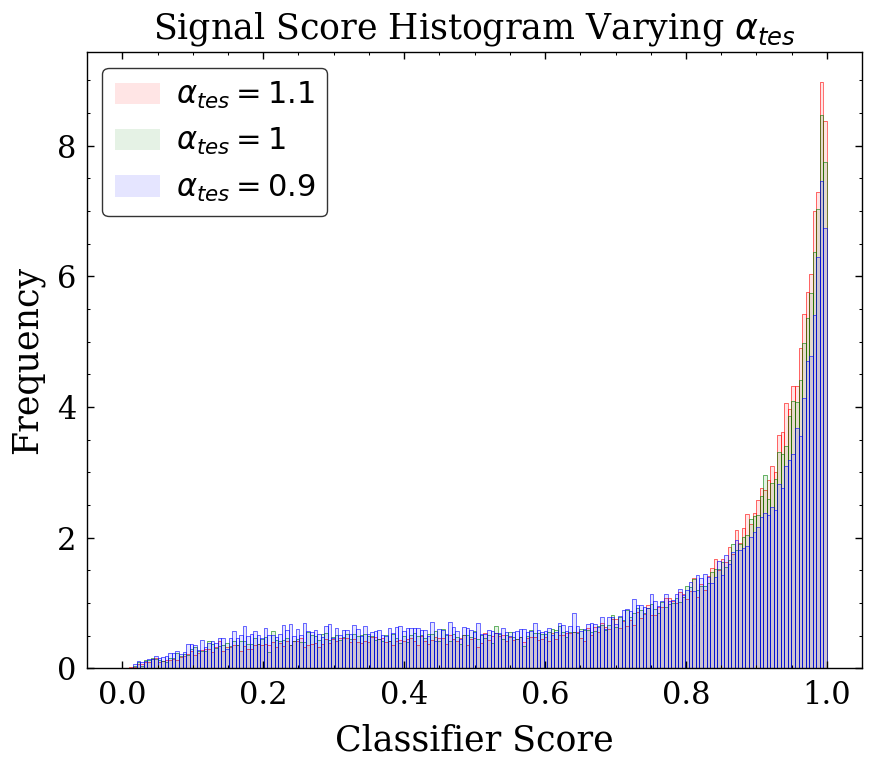}
    \label{fig:signal_score}
  \end{subfigure}
  \begin{subfigure}[t]{0.29\linewidth}
    \centering
    \includegraphics[width=\linewidth]{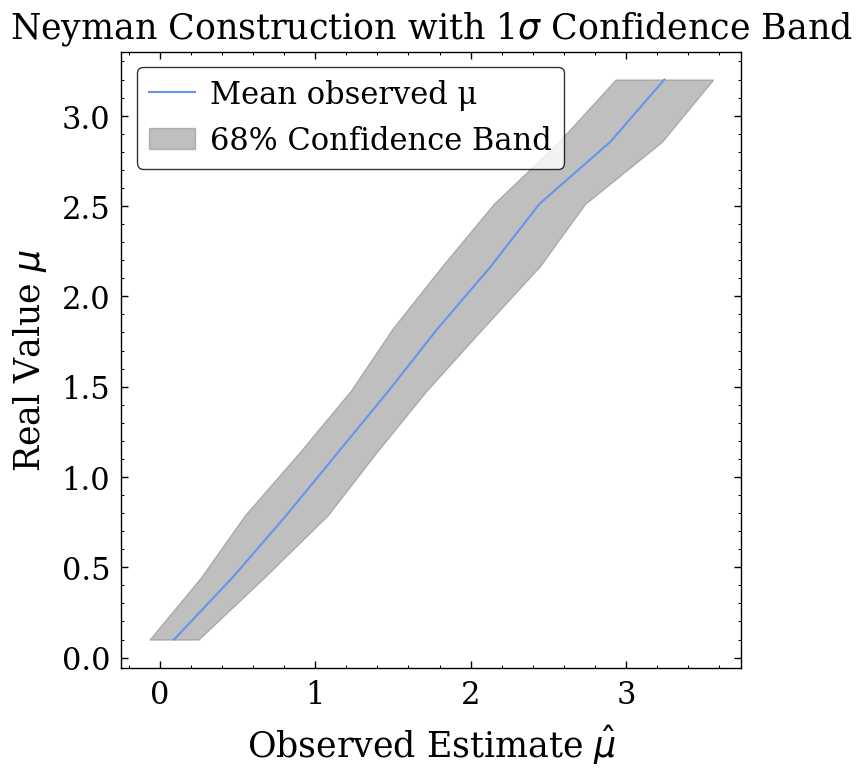}
    \label{fig:neyman}
  \end{subfigure}%
  \caption{CNF distributions for various $c$ (left), DNN score histograms for signal varying the nuisance parameter $\alpha_{\rm tes}$ (center panel), the Neyman confidence belt (right)}
  \label{fig:HiggsMLmethod}
\end{figure}

\textbf{Contrastive Normalising Flows (CNFs)}
A CNF is a normalising flow trained with a \emph{contrastive} objective that simultaneously \textit{maximises} the likelihood of one class and \textit{suppresses} the likelihood of the other. Starting from the standard NF loss, and training on labelled data $\mathbf{x}_s \sim p_s$ and $\mathbf{x}_b \sim p_b$, we insert a term
$c\,\log p_\theta^{(s)}(\mathbf{x}_b)$ so that
\begin{equation}
\begin{split}
\label{eq:cnfloss}
        &\mathcal{L}_s = \frac{1}{|\mathcal{D}|} \sum_{\mathbf{x}_s,\mathbf{x}_b\in\mathcal{D}} \Big\{-\log p_\theta^{(s)}(\mathbf{x}_s) +c\,\log p_\theta^{(s)}(\mathbf{x}_b) \Big\}
\bigg\}
\end{split}
\end{equation}
thereby causing the learned density
$p_\theta^{(s)}$ to concentrate probability mass in regions characteristic of the signal and unlike background. CNFs have been used in anomaly detection settings \cite{schmier2023positivedifferencedistributionimage}. We generalize with $c$ and develop a novel architecture and training procedure empirically required for accurate learning \cite{elsharkawy2025contrastivenormalizingflowsuncertaintyaware}.  Exchanging the roles of $\mathbf{x}_s$ and $\mathbf{x}_b$ gives a loss function $\mathcal{L}_b$ and a learned function $p_\theta^{(b)}$ that concentrates in background regions. Transforming these probabilities as $\Phi_{s,b}(\mathbf{x})=p_\theta^{(s,b)}(\mathbf{x})/\bigl[1+p_\theta^{(s,b)}(\mathbf{x})\bigr]$
gives us our monotonic discrimination functions that retain the full shape of each class. Because the model learns a class distribution, not just a decision boundary, its scores are more stable under systematic shifts than those of a purely discriminative network. Tuning $c$ lets us trade off coverage versus stability under systematic shifts seen in \autoref{fig:HiggsMLmethod}.

\paragraph{The Method} The total training time is 10 GPU hours.

\textbf{Step 1. Pre‑processing.}
Events are split into 1‑jet and 2‑jet categories (empirically, 0‑jet events hurt performance).  
We take the log of features which peak near zero and then standardise all features.

\textbf{Step 2. CNF density learning.}
For each jet category we fit \emph{two} CNFs $\bigl(p^{(s)}_{\theta,c},\,p^{(b)}_{\theta,c}\bigr)$ for $
  c\in\{0.5,\,2.0\}.$ $c>1$ sharpens
signal‑rich regions and is empirically shift‑robust, while $c<1$ preserves coverage. 

\textbf{Step 3. DNN Classifier}
For any event $\mathbf{x}$ we compute $\Phi^{(s,b)}(\mathbf{x})=\frac{p^{(s,b)}_{\theta,c}(\mathbf{x})}{1+p^{(s,b)}_{\theta,c}(\mathbf{x})}$ for $
  c\in\{0.5,\,2.0\}$ yielding four CNF scores per jet category.  Together with the primary and derived features, these are fed to a two‑headed DNN (shared trunk, jet‑specific heads) whose binary‑cross‑entropy loss is minimised on just 1,000 \emph{shifted} mixtures uniformly sampling each $\nu_i$. We highlight the efficacy of CNF features with the relative invariance of the score histogram in \autoref{fig:HiggsMLmethod}.
  
  \textbf{Step 4. Maximum likelihood estimation and the Neyman Construction.}
After training, the classifier scores are histogrammed for a given test set, and maximum likelihood estimation is performed to find point estimates for $\mu$, $\alpha_{jes}$, and $\alpha_{tes}$ given spline-interpolated signal and background template histograms. The point estimate for $\mu$, $\hat{\mu}$, is used to build a Neyman confidence belt, where for each value of real $\mu$ we estimate $\hat{\mu}$ and compute the 68\% spread as can be seen in \autoref{fig:HiggsMLmethod}. This confidence belt can then be inverted at evaluation time to find the 1$\sigma$ error bars on $\mu$ given a $\hat{\mu}$ value.

%% file: hzume.tex
\subsection{\texttt{hzume}: Decision‑Tree Aggregated Features and Hybrid Bin‑Classifier/Quantile‑Regressor}
\label{subsec:hzume}
We build a two‑stage model composed of an \emph{Aggregation stage}
and an \emph{Estimation stage}. 
Total training time is one CPU core hour.

{\bf Aggregation Stage: Feature Engineering} : (i) For each event $(x_{ij})$ a decision tree estimates the class label
        $y_{ij}$ (signal vs.\ background), yielding a probability
        $p_{ij}$.  From the set $\{p_{ij}\}$ we compute and aggregate
        mean, variance, skewness, kurtosis, and the
        empirical quantiles at levels $0$–$255$. 
        (ii) For each feature $x_{ij}$ its mean and variance across events, is fed into a second decision tree that predicts the
        Nuisance Parameters (e.g.\ TES, JES).  These predictions are appended as
        additional features. 

{\bf Estimation Stage: Two Models \& Merging Strategy}  (i) A decision tree classifier partitions the interval $[0.1, 3]$ into five equal‑width bins and predicts the bin containing $\mu$.  The resulting
        probability is converted into the narrowest CI
        covering $68\,\%$ of the total probability.
        (ii) A   quantile‑regression model directly predicts the lower and
        upper quantiles, providing an alternative CI for $\mu$.

\textbf{Model Selection Rule.}  Empirically, the quantile regressor loses accuracy when $\mu$ is near the
end‑points ($0.1$ or $3$).  Therefore, we adopt the bin classifier in the edge regions and the quantile regressor in the central region to produce the final
CI.

%% file: subplots_PRI.tex
\begin{figure}
    \centering
    \includegraphics[width=1\linewidth]{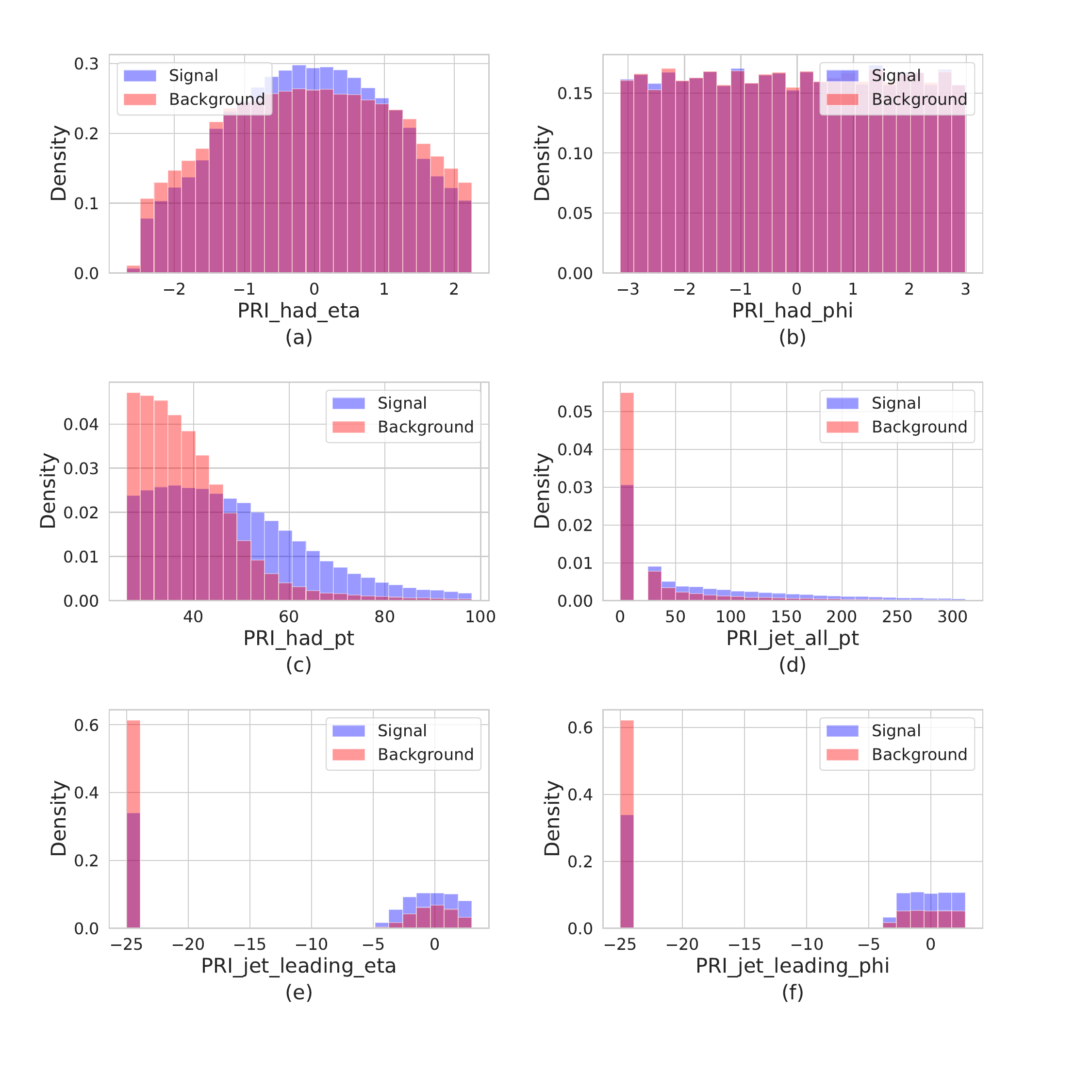}
    \caption{Distributions of: (a) hadron $\eta$, (b) hadron $\phi$, (c) hadron $p_T$, (d) all jets $p_T$, (e) leading jet $\eta$, and (f) leading jet $\phi$. For jet quantities, the left most bin is the default value in the absence of jets.}    
    \label{fig:PRI-1}
\end{figure}

\begin{figure}
    \centering
    \includegraphics[width=1\linewidth]{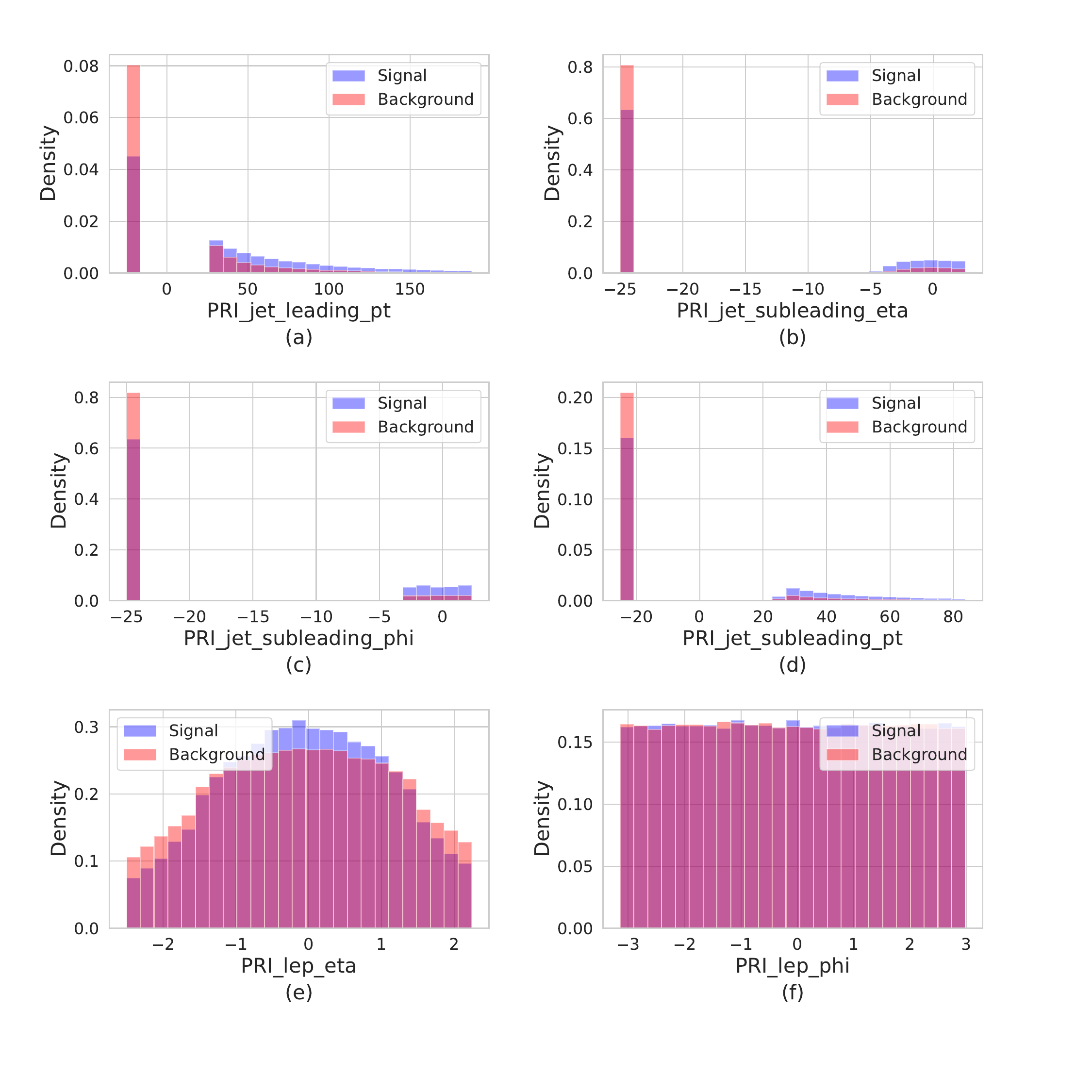}
    \caption{Distributions of: (a) leading jet $p_T$, (b) subleading jet $\eta$, (c) subleading jet $\phi$, (d) subleading jet $p_T$, (e) lepton $\eta$, and (f) lepton $\phi$. For jet quantities, the left most bin is the default value in no jet, or only one jet.}    
    \label{fig:PRI-2}
\end{figure}

\begin{figure}
    \centering
    \includegraphics[width=1\linewidth]{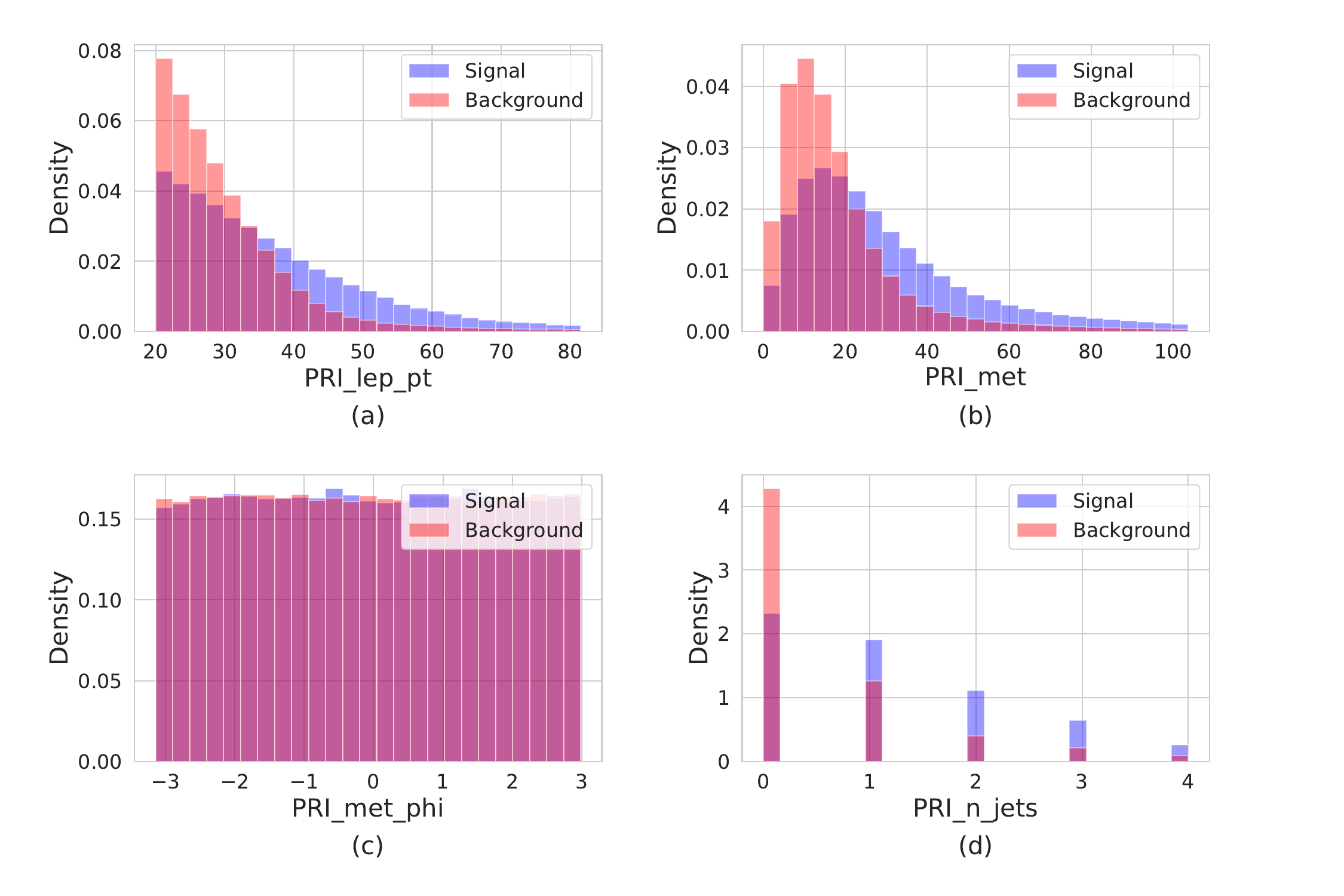}
    \caption{Distributions of: (a) lepton $p_T$, (b) MET, (c) MET $\phi$, and (d) number of jets.}
    \label{fig:PRI-3}
\end{figure}

%% file: subplots_DER.tex
\begin{figure}
    \centering
    \includegraphics[width=1\linewidth]{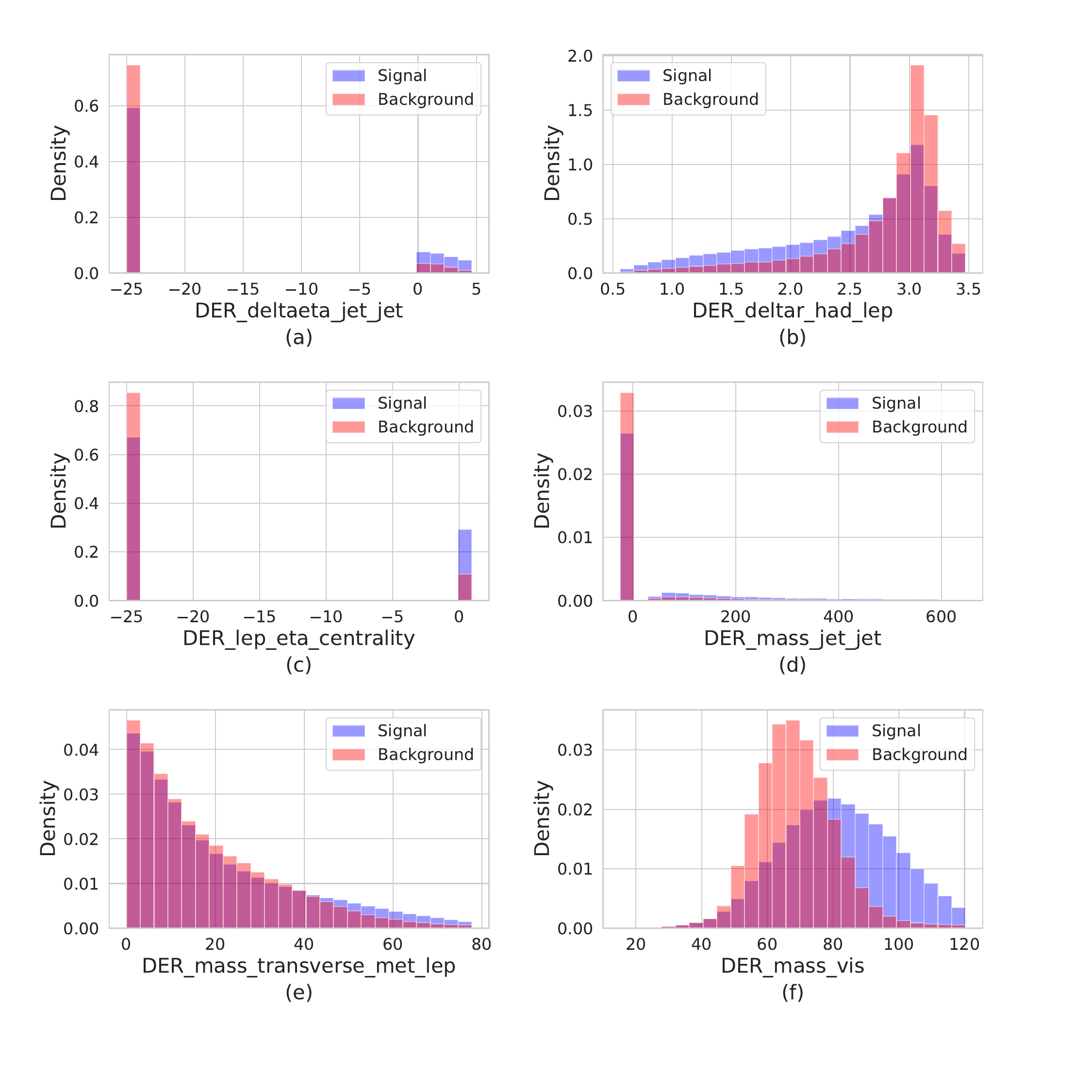}
    \caption{Distributions of kinematic variables: (a) $\Delta\eta(jet\text{-}jet)$, (b) $\Delta R(had\text{-}lep)$, (c) $lep\ \eta$ centrality, (d) $m(jet\text{-}jet)$, (e) $m_T(\text{MET}\text{-}lep)$, and (f) visible mass.}
    \label{fig:DER-1}
\end{figure}

\begin{figure}
    \centering
    \includegraphics[width=1\linewidth]{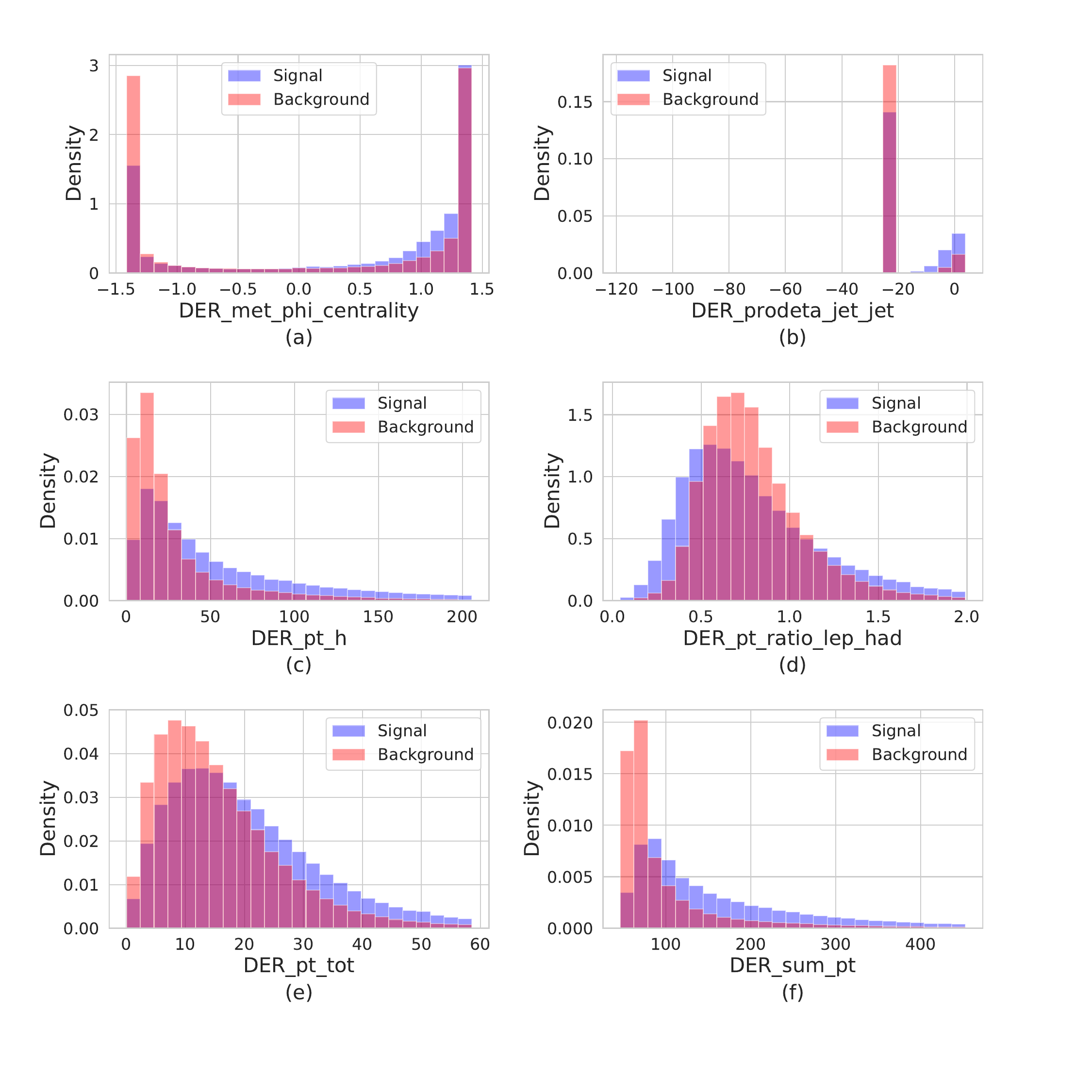}
    \caption{Distributions of: (a) MET $\phi$ centrality, (b) ${\rm prod\ }\eta(jet\text{-}jet)$, (c) $p_T^h$, (d) $p_T(lep\text{/}had)$ ratio, (e) $p_T^{\text{tot}}$, and (f) $\sum p_T$.}
    \label{fig:DER-2}
\end{figure}